\documentclass[12pt]{article}

\usepackage{authblk}
\usepackage{blindtext}

\usepackage{amsmath}
\usepackage{graphicx,psfrag,epsf}
\usepackage{enumerate}
\usepackage{url} 

\usepackage{algpseudocode,algorithm}

\usepackage{amsmath,amsfonts,amssymb,amsthm,amsxtra,bbm,upref,graphicx,epsfig,mathrsfs,tabularx,bm,times,color, multirow}
\usepackage{verbatim}
\usepackage{booktabs}
\usepackage{caption}
\usepackage{graphicx}
\usepackage{subcaption}
\usepackage{comment}
\usepackage[font=footnotesize,labelformat=simple]{subcaption}
\usepackage[numbers]{natbib}

\usepackage[printwatermark]{xwatermark}
\usepackage{xcolor}

\usepackage{graphicx}

\usepackage{url}
\usepackage{verbatim}
\usepackage{siunitx} 

\usepackage{rotating} 
\usepackage{csquotes} 
\usepackage{epigraph}

\usepackage{recycle} 

\usepackage{soul} 

\newcommand{\be}{\begin{equation}}
	\newcommand{\enq}{\end{equation}}
\renewcommand{\baselinestretch}{1}
\makeatletter
\newcommand*{\indep}{%
	\mathbin{%
		\mathpalette{\@indep}{}%
	}%
}

\usepackage{epstopdf}
\AppendGraphicsExtensions{.tif}

\date{} 

\begin{document}

\def\spacingset#1{\renewcommand{\baselinestretch}
{#1}\small\normalsize} \spacingset{1}




\title{\bf Proactive Anomaly Screen for Multiple Endpoints Using Bayesian Latent Class Modeling: A $k$-Step Ahead Approach}

\author[1]{Yuxi Zhao \thanks{Corresponding Author: Yuxi.Zhao@pfizer.com}}
\author[1]{Margaret Gamalo}
\affil[1]{ Inflammation, Immunology \& Specialty Care Statistics, Pfizer Inc}
\maketitle

\bigskip

\begin{abstract}
In clinical trials, ensuring the quality and validity of data for downstream analysis and results is paramount, thus necessitating thorough data monitoring. This typically involves employing edit checks and manual queries during data collection. Edit checks consist of straightforward schemes programmed into relational databases, though they lack the capacity to assess data intelligently. In contrast, manual queries are initiated by data managers who manually scrutinize the collected data, identifying discrepancies needing clarification or correction. Manual queries pose significant challenges, particularly when dealing with large-scale data in late-phase clinical trials. Moreover, they are reactive rather than predictive, meaning they address issues after they arise rather than preemptively preventing errors. In this paper, we propose a joint model for multiple endpoints, focusing on primary and key secondary measures, using a Bayesian latent class approach. This model incorporates adjustments for risk monitoring factors, enabling proactive, $k$-step ahead, detection of conflicting or anomalous patterns within the data. 
Furthermore, we develop individualized dynamic predictions at consecutive time-points to identify potential anomalous values based on observed data. This analysis can be integrated into electronic data capture systems to provide objective alerts to stakeholders. We present simulation results and demonstrate effectiveness of this approach with real-world data.
\end{abstract}

\noindent%
\vfill

\spacingset{1.45} 

\section{Introduction}
High-quality data is essential for ensuring the reliability, accuracy, and integrity of trial results, thereby significantly impacting the credibility and confidence in study outcomes \citep{al2005effect}. Achieving quality data in clinical trials requires monitoring data that needs a comprehensive allocation of resources spanning personnel, technology, training, and adherence to regulatory standards. In the personnel aspect, collaboration among various stakeholders, including clinicians, data managers, statisticians, site personnel, and clinical research auditors is necessary. Compliance entails adherence to extensive guidance and widely accepted practices aimed at standardizing the data collection process and quantifying data quality (\cite{molloy2016monitoring}). Guidance such as the Good Clinical Data Management Practices (GCDMP) provided by the Society for Clinical Data Management (SCDM) offers comprehensive direction on all aspects related to data management across the stages of clinical trials.

Monitoring strategies in clinical trials can be categorized into three approaches: extensive, reduced, and targeted. Extensive monitoring involves 100\% source data verification (SDV) for primary and key secondary endpoints. Reduced monitoring employs verification through a risk-adapted approach or random sampling of centers, patients, and outcomes. Targeted monitoring utilizes key risk indicators and statistical monitoring to identify potential issues or anomalies in the data. {\it Risk-based monitoring} (RBM) aims for real-time remediation of potential errors based on critical risk assessments. This approach contrasts with passive monitoring of past events and focuses on addressing issues that could impact participant safety and data quality. The majority of data monitoring tools and methodologies within the scope of RBM primarily concentrate on overseeing and managing data entry errors and alterations. These include systems such as Target e*CRF \cite{mitchel2011}. Additionally, a quality review conducted by \citet{tolmie2011clinical} offered descriptive and comparative analyses of data queries categorized by the country of origin. 

A strategic approach to data analysis can be implemented via {\it statistical monitoring} (StM),  as suggested by \citet{baigent2008ensuring}. Humans often generate poor random sequences, so checking for randomness (e.g., using Benford’s Law on first digits or digit preference) is insightful. Fabricated data is often less plausible than real data, making it effective to examine plausibility through correlation structures, outliers, inliers, and dates. Clinical trial data is highly structured, allowing for the assessment of comparability across different centers or treatment arms.  \citet{van2005data} generalize data cleaning as a process from data screening to documentation and reporting. They identify four types of anomalies for error screening: insufficient or excessive data, outliers (including consistencies), unusual patterns in distributions (both singular and joint), and unexpected analysis outcomes. The authors recommend using descriptive statistical techniques (e.g., \citet{ bauer2000editing, epiinfo, epidata}) to identify questionable data patterns, including inliers and outliers. However, these traditional methods lack automated detection and statistical learning capabilities. Identifying anomalous data requires comprehensive information about the data distribution, including normal ranges, distribution shapes, strength of associations, and temporal patterns. This knowledge is crucial for effectively distinguishing between valid data and anomalies.

In this paper, we propose a joint model for multiple endpoints (primary and key secondary) using a Bayesian latent class approach adjusted for risk monitoring factors. This method leverages latent classes to capture heterogeneity and detect anomalies deviating from expected patterns. The flexibility of Bayesian methods allows for individualized dynamic predictions ($k$-step ahead), crucial for identifying potential anomalies over time. As new data becomes available, Bayesian models can adapt and update predictions, providing a real-time monitoring system. Embedding this analysis in electronic data capture systems enables seamless integration into clinical workflows, allowing timely alerts to clinicians and patients. This proactive approach ensures prompt identification and resolution of potential issues, enhancing data quality and patient safety in clinical trials.

The paper follows a structured outline: Section 2 outlines the proposed approach and details the derivation of individualized dynamic prediction. Section 3 delves into the simulation settings and presents the results obtained. In Section 4, the focus shifts to real-world applications, providing insights into how the proposed approach can be applied in practical scenarios. Finally, Section 5 offers a discussion of the findings and outlines potential avenues for future research.

\section{Methodology}

Typically, data from clinical sites is transferred to a centralized database at regular intervals, often weekly or monthly, enabling monitoring and analysis at regular intervals. However, in certain instances, data transfer may occur more frequently, especially in trials involving critical safety endpoints or where real-time or near-real-time monitoring is feasible, such as with electronic data capture systems. 

\subsection{Illustrative Example}
\begin{figure}
    \centering
    \includegraphics[width=0.75\linewidth]{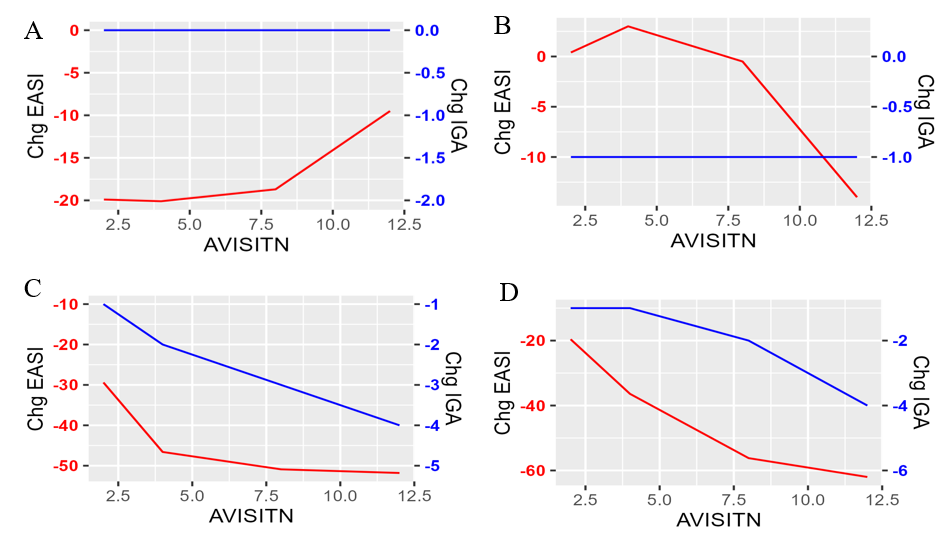}
    \caption{An Illustrative Example: Deviating Patterns between multiple endpoints of Change from baseline in EASI ("Chg in EASI" denoted in red) and Change from baseline in IGA ("Chg in IGA" denoted in blue). }
    \label{fig:example}
\end{figure}

Atopic dermatitis (AD), a prevalent condition often originating in early childhood, is commonly assessed for efficacy using endpoints such as the Eczema Area and Severity Index (EASI) score,
Investigator’s Global Assessment (IGA) score, and Scoring Atopic Dermatitis (SCORAD) index. We will use EASI and IGA for an illustration with measurements at Week 2, 4, 8 and 12 using change from baseline. As shown in Figure \ref{fig:example}, it is common to observe deviating patterns between multiple efficacy endpoints, which could signal data errors or issues. As shown in the figure of four individual plots,  A and B are easy cases for identification of deviating patterns; C and D are the subtle cases for eyeballing, where the slope of changing is deviating between the endpoints.

To ensure data quality, a statistician plays a critical role in monitoring ongoing prospective study data to identify unusual or clustered patterns that may signal underlying issues. This process involves designing and implementing edit checks and validation procedures to flag errors or inconsistencies in the collected data. Traditionally, many of these edit checks are static, relying on fixed numerical thresholds at each visit, often derived from clinically plausible values in the literature.

\begin{figure}
    \centering
    \includegraphics[width=0.8\linewidth]{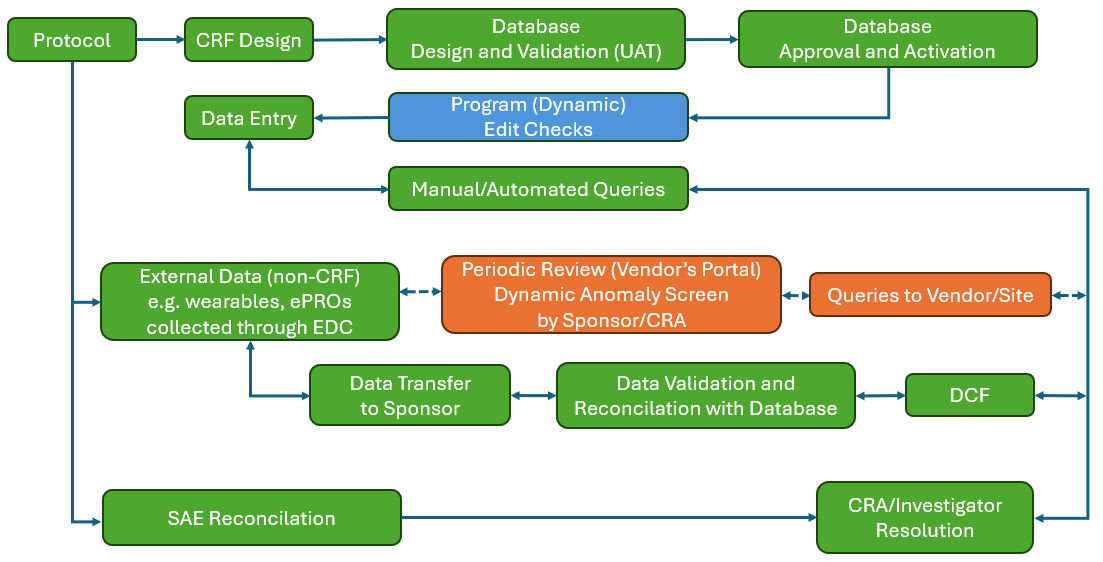}
    \caption{Clinical data management workflow with proposal of embedded workflow (in blue and orange).}
    \label{fig:DMprocess}
\end{figure}

In contrast, we propose a dynamic approach to edit checks and external data especially efficacy data through periodical reviews that incorporates temporal patterns in the accumulating data, enabling earlier screening of anomalies. This proactive method allows investigators to identify potential issues before they finalize data and avoid submitting Data Clarification Forms (Figure ~\ref{fig:DMprocess}) which can be costly. Early screening is particularly valuable because modifying data after it has been finalized in an electronic capture system can be challenging, as finalized data often becomes the source data. By leveraging dynamic checks, this method enhances the reliability and integrity of study data in real time.

\subsection{Bayesian latent class approach}

Suppose we have two longitudinal efficacy endpoints capturing change from baseline for participant $i$ at times $t_{i,j}^x$ and $t_{i,j}^y$, denoted as $x_i := \{x_{i,j}:j=1\cdots, n_i^x\}$ and $y_i=\{y_{i,j}:j=1,\cdots,n_i^y\}$ for $i=1,\cdots,n$. Let $C$ be the maximum number of latent classes and $z_{i}$ represent the latent class membership of participant $i$. Assuming conditional independence given the latent class membership, the two longitudinal processes are modeled as:
\begin{align}
x_{i,j} (t_{i,j}^x)\mid \{z_{i}=c\} =& \beta_{0c}^x  +  \beta_{0,base}^x Baseline^x  +
\beta_{1c}^x t_{i,j}^x + \beta_{2c}^x (t_{i,j}^x)^2 +  v_{s(i)}^x + w_i^x + e_{i,j}^x \label{eq:modelSetup1}\\
y_{i,j} (t_{i,j}^y)\mid \{z_{i}=c\} =& \beta_{0c}^y  + \beta_{0,base}^y Baseline^y  + \beta_{1c}^y  t_{i,j}^y + \beta_{2c}^y (t_{i,j}^y)^2 + v_{s(i)}^y + w_i^y + e_{i,j}^y \label{eq:modelSetup2}
\end{align}
where $v_{s(i)}^x \sim N(0, 1/\tau_{sc}^x ),v_{s(i)}^y \sim N(0, 1/\tau_{sc}^y)$ are class specific errors, $w_i^x \sim N(0, 1/\tau_w^x), $ and $w_i^y \sim N(0,1/\tau_w^y)$ are errors associated to the outcome measure, and $ e_{i,j}^x\sim N(0,1/\tau_e^x)$ and $ e_{i,j}^y\sim N(0,1/\tau_e^y)$ are common errors. 

In this context, each observation is assumed to originate from one of many heterogeneous processes, with those arising from infrequently occurring processes considered anomalies. This concept can be addressed through clustering or partitioning, which aims to distinguish between observations generated by different longitudinal processes and identify significant deviations. The Dirichlet process mixture (DPM) model has become popular due to its ability to cluster similar observations. Previous research by Quintana and colleagues has highlighted the utility of DPM models in detecting anomalous data and developed a sophisticated decision theory to balance model complexity and optimal parameter estimation \citep{quintana2003bayesian, quintana2022dependent}.

Assume we have discrete monitoring factors, such as sites, totalling $M$. For each subject $i$, the indicator vector $s_i\in \{1,\cdots,M\}$ represents site membership. To incorporate this into the Dirichlet prior, we can define a data generation process with a non-informative prior as follows:
\begin{align*}
s_i \mid \{z_{i}=c\} \sim &\ \text{Categorical}(p_{1c},\cdots,p_{Mc})\\
p_{1c},\cdots,p_{Mc} \sim &\ \text{Dirichlet}(1,\cdots,1).
\end{align*} Here, the categorical distribution assigns probabilities to the site memberships for each latent class $c$, while the Dirichlet distribution provides a prior over these probabilities with each
where $p_{1c} = \cdots = p_{Mc} = 1$,  indicating a non-informative prior. The prior for the assignment of latent classes is based on the symmetric Dirichlet approximation, as discussed by \citet{neal2000markov, rasmussen1999infinite}. \citet{li2019tutorial} provides technical details and mathematics on this DPM setup. It can be efficiently implemented using probabilistic programming languages like JAGS, which facilitates fast mixing for Markov Chain Monte Carlo (MCMC) methods. Assume $C$ is an arbitrarily large number representing the potential latent classes: 
\begin{align*}
z_i\sim &\ \text{Categorical}(\pi_1,\cdots,\pi_C)\\
\pi_1,\cdots,\pi_C \sim &\ \text{Dirichlet}(\alpha/C,\cdots,\alpha/C).
\end{align*} In this context, $\alpha$ serves as the precision parameter in the DPM. The conditional probability of assigning subject $i$ to cluster $c$ is given by $\text{Pr}(z_i=c\mid z_{-i},\alpha)=\frac{n_{-i,c}+\alpha/C}{n-1+\alpha}$ \citep{li2019tutorial}, where $z_{-i}$ denotes the cluster memberships of all subjects except the $i$-th, and $n_{c,-i}$ is the size of cluster $c$ excluding subject $i$. Given a relatively large $C$, the membership probability is mostly affected by the ratio of $n_{-i,c}$ over $n$. The DPM model induces a prior distribution over the data partition, enabling inference on the statistics of the data partition, such as the number of clusters, based on the posterior distribution. By manipulating the Dirichlet process precision parameter, one can derive a simple and intuitive criterion for detecting anomalous data with DPM models. To provide further context, the prior distributions for the model parameters are specified as non-informative conjugate priors for each class $c=1,\cdots,C$, the details can be found in the appendix.

With the posterior distributions of these parameters, we can establish posterior predictions for future observations based on individual observed data by marginalizing the posterior distribution conditioned on the latent classes. This allows us to derive credible intervals from the posterior predictions to enable anomaly detection for each individual. After accumulating appropriate sized of data, an edit check can be set up with pre-trained results of the data accumulated so far.  At each monitoring or review timepoint in the data management process, these pre-trained results can be updated with newly accumulated data. The following scenarios illustrate posterior prediction implementation.

\subsection{Posterior Prediction}

\subsubsection{Case 1}

If we have newly observed data, we can condition on the previously accumulated data (historical data) to obtain the posterior probability of the newly observed data. Say for subject $i$, we have sequential newly observed data at times $t_{i,1}^\star$, $t_{i,2}^\star$, $t_{i,3}^\star$, and the observed data given by $x_{i,t_{i,1}}^\star,x_{i,t_{i,2}}^\star,x_{i,t_{i,3}}^\star$ and $y_{i,t_{i,1}}^\star,y_{i,t_{i,2}}^\star,y_{i,t_{i,3}}^\star$. The posterior distribution for these data is
\begin{equation}
x_{i,t_{i,1}}^\star,x_{i,t_{i,2}}^\star,x_{i,t_{i,3}}^\star, y_{i,t_{i,1}}^\star,y_{i,t_{i,2}}^\star,y_{i,t_{i,3}}^\star \mid  \textbf{Data}
\end{equation}
This posterior distribution is obtained by integrating over the parameter space of $\theta$
\begin{equation}
 \int f(x_{i,t_{i,1}}^\star,x_{i,t_{i,2}}^\star,x_{i,t_{i,3}}^\star, y_{i,t_{i,1}}^\star,y_{i,t_{i,2}}^\star,y_{i,t_{i,3}}^\star \mid  \theta) \text{Pr}(\theta\mid\textbf{Data}) d\theta
\end{equation} Where $f(\cdot)$ is the likelihood function and $Pr(\theta\mid\textbf{Data})$ is the posterior distribution of the parameters given accumulated data. By assuming a finite mixture model with $C$ latent classes, this can be further detailed as:
\begin{equation}
 \int  \sum_c f\left(x_{i,t_{i,1}}^\star,x_{i,t_{i,2}}^\star,x_{i,t_{i,3}}^\star \mid \theta_c \right) f\left(y_{i,t_{i,1}}^\star,y_{i,t_{i,2}}^\star,y_{i,t_{i,3}}^\star \mid \theta_c \right)\text{Pr}(z_{i}=c\mid \theta) \times\text{Pr}(\theta\mid\textbf{Data}) d\theta.
\end{equation} Here, $\theta$ includes all model parameters $\{\beta_{0,\text{base}}^x,\beta_{0,\text{base}}^y,\tau_w^x,\tau_w^y, \tau_e^x,\tau_e^y\beta_{0c}^x,\beta_{1c}^x,\beta_{2c}^x,\tau_{sc}^x,\beta_{0c}^y, \beta_{1c}^y,\beta_{2c}^y, \tau_{sc}^y,$ $ p_{1c},\cdots, p_{Mc},\pi_1,\cdots,\pi_C , c=1,\cdots,C \}$. The likelihood $f\left(x_{i,t_{i,1}}^\star,x_{i,t_{i,2}}^\star,x_{i,t_{i,3}}^\star \mid \theta_c \right)$ is multivariate normal 
\begin{equation*} 
\begin{pmatrix}  x_{i,t_{i,1}}^\star \\
x_{i,t_{i,2}}^\star\\ x_{i,t_{i,3}}^\star\\ \end{pmatrix} \mid \theta_c^{(m)} \sim \text{MVN}(\mathbf{\mu}_{c,x^{\star}},\mathbf{\Sigma}_{c,x^{\star}}),
\end{equation*} where the mean vector $\mu_{c}^{x^*}$ is
\begin{equation*}
 \mu_{c,x^{\star}} = \begin{pmatrix}
\beta_{0c}^x  +  \beta_{0,\text{base}}^x \text{Baseline}^x  +
\beta_{1c}^x t_{i,1}^\star + \beta_{2c}^x (t_{i,1}^\star)^2\\
\beta_{0c}^x  +  \beta_{0,\text{base}}^x \text{Baseline}^x  +
\beta_{1c}^x t_{i,2}^\star + \beta_{2c}^x (t_{i,2}^\star)^2\\
\beta_{0c}^x  +  \beta_{0,\text{base}}^x \text{Baseline}^x  +
\beta_{1c}^x t_{i,3}^\star + \beta_{2c}^x (t_{i,3}^\star)^2\\
\end{pmatrix},
\end{equation*} and the covariance matrix $\mathbf{\Sigma}_c$ is
\begin{equation*}
\mathbf{\Sigma}_{c,x^{\star}}=\begin{pmatrix} 1/\tau_{sc}^x +1/\tau_w^x +1/\tau_e^x &1/\tau_{sc}^x+1/\tau_w^x &1/\tau_{sc}^x+1/\tau_w^x\\
1/\tau_{sc}^x+1/\tau_w^x & 1/\tau_{sc}^x+1/\tau_w^x+1/\tau_e^x & 1/\tau_{sc}^x+1/\tau_w^x\\
1/\tau_{sc}^x+1/\tau_w^x & 1/\tau_{sc}^x+1/\tau_w^x & 1/\tau_{sc}^x+1/\tau_w^x+1/\tau_e^x \end{pmatrix}.
\end{equation*} The likelihood $f\left(y_{i,t_{i,1}}^\star,y_{i,t_{i,2}}^\star,y_{i,t_{i,3}}^\star \mid \theta_c \right)$ is defined similarly. Built on these, the posterior prediction for the future consecutive data can be approximated by 
\begin{equation*}
    \frac{1}{Q}\sum_{q=1}^Q  \sum_c f\left(x_{i,t_{i,1}}^\star,x_{i,t_{i,2}}^\star,x_{i,t_{i,3}}^\star\mid \theta_c^{(q)} \right) f\left(y_{i,t_{i,1}}^\star,y_{i,t_{i,2}}^\star,y_{i,t_{i,3}}^\star \mid \theta_c^{(q)} \right) \pi_c^{(q)}
\end{equation*} where $Q$ is an arbitrarily large number. 

For the computation of multivariate normal distribution with compound symmetry as covariance structure, we can apply a transformation to the vector and obtain the likelihood from the product of univariate normal distribution. For simplicity, we denote $\Sigma_c$ as $\Sigma_{c,x^{\star}}$ in the following derivation. Given $\Sigma_c^{x^*} \in \mathbb{R}^{n_i^x}$, it can be rewritten as  
\begin{align*}
    \Sigma_c = \widetilde{\sigma}_c^2\left[\text{diag}(1-\rho_c,\cdots, 1-\rho_c)+\rho_c 11^T\right] :=\widetilde{\sigma}_c^2 \widetilde{\Sigma}_c
\end{align*}
with $\widetilde{\sigma}_c^2=1/\tau_{sc}^x +1/\tau_w^x+1/\tau_e^x$ and $\rho_c=(1/\tau_{sc}^x+1/\tau_w^x)/(1/\tau_{sc}^x+1/\tau_w^x+1/\tau_e^x)$. Define $\widetilde{\Sigma}_c=PD_cP^T$ by eigenvalue decomposition with two eigenvalues $1+(n_i^x-1)\rho_c$ of multiplicity 1 and $1-\rho_c$ of multiplicity $n_i^x-1$; note that $P$ can be precomputed with regard to the dimension $n_i^x$ independent of $\rho_c$.  In order to save computation time, we calculate $P^T x_{i}^\star \sim N(P^T\mu_{c}, \widetilde{\sigma}_c^2 D_c)$ with the first element $[P^T x_{i}^\star]_1\sim N([P^T\mu_{c}]_1,\widetilde{\sigma}_c^2(1+(n_i^x-1)\rho_c))$ and the remaining elements $[P^T x_{i}^\star]_u\sim N([P^T\mu_{c}]_u,\widetilde{\sigma}_c^2(1-\rho_c))$ for $u=2,\cdots,n_i^x$.

\subsubsection{Case 2}

If we have a fitted model and posterior samples from historical or accumulated data, we can use them to predict future trends, given the newly observed/accumulated data vectors ${[x_{i,t_{i,j}}^\star]}$ and ${[y_{i,t_{i,j}}^\star]}$ of an individual. For a subject $
i$ with observed time points up to $t_{i,j}^\star$ for both endpoints, we can predict the trend for future time points like $t_{i,j+1}^\star$, $t_{i,j+2}^\star$, $t_{i,j+3}^\star$. The posterior distribution for these future data points, conditioned on the observed data and the accumulated data, is given by:
\begin{align*}
&x_{i,t_{i,j+1}}^\star,x_{i,t_{i,j+2}}^\star,x_{i,t_{i,j+3}}^\star, y_{i,t_{i,j+1}}^\star,y_{i,t_{i,j+2}}^\star,y_{i,t_{i,j+3}}^\star \mid {[x_{i,t_{i,j}}^\star]}, {[y_{i,t_{i,j}}^\star]}, \textbf{Data}\\
=& \int f(x_{i,t_{i,j+1}}^\star,x_{i,t_{i,j+2}}^\star,x_{i,t_{i,j+3}}^\star, y_{i,t_{i,j+1}}^\star,y_{i,t_{i,j+2}}^\star,y_{i,t_{i,j+3}}^\star \mid {[x_{i,t_{i,j}}^\star]}, {[y_{i,t_{i,j}}^\star]}, \theta) \text{Pr}(\theta\mid\textbf{Data}) d\theta\\
=& \int \frac{ f\left(x_{i,t_{i,j+1}}^\star,x_{i,t_{i,j+2}}^\star,x_{i,t_{i,j+3}}^\star, y_{i,t_{i,j+1}}^\star,y_{i,t_{i,j+2}}^\star,y_{i,t_{i,j+3}}^\star, [x_{i,t_{i,j}}^\star], [y_{i,t_{i,j}}^\star] \mid \theta \right) }{f\left([x_{i,t_{i,j}}^\star], [y_{i,t_{i,j}}^\star] \mid \theta \right)} \text{Pr}(\theta\mid\textbf{Data}) d\theta\\
=& \int \frac{ \sum_c f\left(x_{i,t_{i,j+1}}^\star,x_{i,t_{i,j+2}}^\star,x_{i,t_{i,j+3}}^\star, [x_{i,t_{i,j}}^\star] \mid \theta_c \right) f\left(y_{i,t_{i,j+1}}^\star,y_{i,t_{i,j+2}}^\star,y_{i,t_{i,j+3}}^\star, [y_{i,t_{i,j}}^\star] \mid \theta_c \right)\text{Pr}(z_{i}=c\mid \theta) }{\sum_c f\left([x_{i,t_{i,j}}^\star] \mid \theta_c \right) f\left([y_{i,t_{i,j}}^\star] \mid \theta_c \right)\text{Pr}(z_{i}=c\mid \theta) } \\
&\quad \times\text{Pr}(\theta\mid\textbf{Data}) d\theta\\
= &\frac{1}{Q}\sum_{q=1}^Q \frac{ \sum_c f\left(x_{i,t_{i,j+1}}^\star,x_{i,t_{i,j+2}}^\star,x_{i,t_{i,j+3}}^\star, [x_{i,t_{i,j}}^\star] \mid \theta_c^{(q)} \right) f\left(y_{i,t_{i,j+1}}^\star,y_{i,t_{i,j+2}}^\star,y_{i,t_{i,j+3}}^\star, [y_{i,t_{i,j}}^\star] \mid \theta_c^{(q)} \right) \pi_c^{(q)} }{\sum_c f\left([x_{i,t_{i,j}}^\star] \mid \theta_c^{(q)} \right) f\left([y_{i,t_{i,j}}^\star] \mid \theta_c^{(q)} \right) \pi_c^{(q)} }
\end{align*}
where $f(\cdot)$ is multivariate normal with $f\left(x_{i,t_{i,j+1}}^\star,x_{i,t_{i,j+2}}^\star,x_{i,t_{i,j+3}}^\star, [x_{i,t_{i,j}}^\star] \mid \theta_c^{(q)} \right) $ defined as below:
\begin{equation*} 
\begin{pmatrix} [x_{i,t_{i,j}}^\star]\\ x_{i,t_{i,j+1}}^\star \\
x_{i,t_{i,j+2}}^\star\\ x_{i,t_{i,j+3}}^\star\\ \end{pmatrix} \mid \theta_c^{(q)} \sim \text{MVN}(\mathbf{\mu}_{c,x^{\star\star}},\mathbf{\Sigma}_{c,x^{\star\star}}), 
\end{equation*} where the mean $\mu_{c,x^{\star\star}}$ is
\begin{equation*}
\mu_{c,x^{\star\star}} = \begin{pmatrix}
\beta_{0c}^x  +  \beta_{0,\text{base}}^x \text{Baseline}^x  +
\beta_{1c}^x t_{i, 1}^\star + \beta_{2c}^x (t_{i,1}^\star)^2\\
\vdots\\
\beta_{0c}^x  +  \beta_{0,\text{base}}^x \text{Baseline}^x  +
\beta_{1c}^x t_{i, j}^\star + \beta_{2c}^x (t_{i,j}^\star)^2\\
\beta_{0c}^x  +  \beta_{0,\text{base}}^x \text{Baseline}^x  +
\beta_{1c}^x t_{i,j+1}^\star + \beta_{2c}^x (t_{i,j+1}^\star)^2\\
\beta_{0c}^x  +  \beta_{0,\text{base}}^x \text{Baseline}^x  +
\beta_{1c}^x t_{i,j+2}^\star + \beta_{2c}^x (t_{i,j+2}^\star)^2\\
\beta_{0c}^x  +  \beta_{0,\text{base}}^x \text{Baseline}^x  +
\beta_{1c}^x t_{i,j+3}^\star + \beta_{2c}^x (t_{i,j+3}^\star)^2\\
\end{pmatrix},
\end{equation*} and the covariance matrix $\mathbf{\Sigma}_{c,x^{\star\star}}$ is
\begin{equation*}
\mathbf{\Sigma}_{c,x^{\star\star}}=\begin{pmatrix} 1/\tau_{sc}^x+ 1/\tau_{w}^x+1/\tau_{e}^x & 1/\tau_{sc}^x+1/\tau_{w}^x &\cdots & 1/\tau_{sc}^x+1/\tau_{w}^x\\
 1/\tau_{sc}^x+1/\tau_{w}^x &  1/\tau_{sc}^x+1/\tau_{w}^x+1/\tau_{e}^x & \cdots & 1/\tau_{sc}^x+1/\tau_{w}^x\\
\vdots & \vdots &\ddots &\vdots\\
 1/\tau_{sc}^x+1/\tau_{w}^x &  1/\tau_{sc}^x+1/\tau_{w}^x &\cdots&  1/\tau_{sc}^x+1/\tau_{w}^x+1/\tau_{e}^x \end{pmatrix}.
\end{equation*}
The likelihoods $f\left(y_{i,t_{i,j+1}}^\star,y_{i,t_{i,j+2}}^\star,y_{i,t_{i,j+3}}^\star, [y_{i,t_{i,j}}^\star] \mid \theta_c^{(m)} \right)$ and $f\left([x_{i,t_{i,j}}^\star] \mid \theta_c^{(m)} \right)$, $f\left([y_{i,t_{i,j}}^\star] \mid \theta_c^{(m)} \right)$ can be defined in a similrly.

If we have certain ranges of interest for $x_{i,t_{i,j+1}}^\star,x_{i,t_{i,j+2}}^\star,x_{i,t_{i,j+3}}^\star, y_{i,t_{i,j+1}}^\star,y_{i,t_{i,j+2}}^\star,y_{i,t_{i,j+3}}^\star$, we can have the posterior probability of the future values fall in the pre-specified region. Also we can find the values with highest posterior density from the pre-specified grid of values.

\subsection{Credible region}

To compute the credible region, we adopt a grid-based approach due to the absence of a closed form solution. This involves creating a fine grid of possible values, denoted as 
$x_{\text{grid}}$ and $y_{\text{grid}}$, to calculate the posterior prediction density $p_{x_\text{grid}, y_\text{grid}}$ for every combination within the grid. Subsequently, we perform resampling from these grid values based on the posterior prediction densities. To simplify computation, we discretize the ranges of endpoint values into regions and compute the posterior prediction probability for each region. By evaluating the posterior prediction probability, we can delineate a credible region, which serves as a criterion to assess the validity of an observation, based on whether it falls within the credible region or not.

Two algorithms are proposed to identify the credible region:
\begin{itemize}
\item {\bf Algorithm 1}: This method involves branching out from the region with the highest density. It comprises three parts: branching out at a faster rate with two steps ahead, branching out at a slower rate with only one step ahead, and tuning on the boundary.
\item {\bf Algorithm 2}: This approach selects the highest density regions, starting from the most dense region, and progressively slices down until achieving the desired coverage probability.
\end{itemize} In {\bf Algorithm 1}, the parameter $c$ determines the rate of branching out. If $c=0$, implying a slow branching out with one step ahead, it might overlook other regions with relatively higher probabilities. Conversely, if $c=\text{Target coverage}$, indicating branching out with two steps ahead, it may include more smaller regions than necessary. Hence, it is advantageous to use a combination of branching rates. In practice, setting $c$ to a constant close to the target coverage generally yields consistent results without significant differences.

\begin{algorithm}[hbt!]
\caption{Branching out}
\begin{algorithmic}[1]
\State Specify a $c < \text{target coverage}$;
\State Start from the highest probability region, denoted as the region index $(u_1^0,v_1^0)$, and save it into the selected region $\mathcal{M}$.
\State Create an empty set to save removed neighbors ($\mathcal{M}_R$) and their corresponding probabilities ($p_R$).
\For {steps $s=1,\cdots$ till $p_\text{sum}>c$ to branch out quickly}
\State Repeat ``Branch out quickly" as shown in \ref{alg:details}
\EndFor
\For {steps $s=1,\cdots$ till $p_\text{sum}>\text{target coverage}$ to slowly branch out and fine tune}
\State Repeat ``Branch out slowly" as shown in \ref{alg:details}
\EndFor
\State Final tuning on the boundary. Select the boundary regions for all $i\in \mathcal{M}$ and save them into a set $\mathcal{D}_1$. Order the corresponding probabilities of $\mathcal{D}_1$ from smallest to largest and rearrange the set accordingly into $\mathcal{D}_1^*$. Compute the cumulative sum of ordered probabilities $p_{1,\text{cum}}^*$ for $\mathcal{D}_1^*$. Remove the first $r$ regions in $\mathcal{M}$ i.e. $\text{max}_r [p_{1,cum}^*]_r<=p_\text{sum}-\text{target coverage}$ where $[\cdot]_r$ denotes $r$th element in the vector. Save the trimmed region to be the final output.
\end{algorithmic}
\end{algorithm}

\begin{algorithm}
\caption{Highest Density Regions}
\begin{algorithmic}[1]
\State Order the posterior prediction probability $(p_1,\cdots,p_M)$ ordered from largest to smallest.
\State Start from the highest probability region, denoted a the region index $(u_1^0,v_1^0)$, and save it into the selected region $\mathcal{M}$.
\For {Step $s = 1,\cdots$ until $p_\text{sum}>\text{target coverage}$ }
        \State Find the regions $(u_i^s,v_i^s)$  with corresponding posterior prediction probability less than $p_s$;
        \State Save the regions $(u_i^s,v_i^s)$ into $\mathcal{M}$ and compute the sum of all the posterior prediction probability $p_\text{sum}$ in  all the selected region set $\mathcal{M}$. 
\EndFor
\end{algorithmic}
\end{algorithm}

%

\subsection{Anomaly Screen}

The primary objective in anomaly screen/detection is to identify observations that significantly deviate from the expected distribution under the assumed model. In our case, an observation is flagged as potentially anomalous if it falls outside a predefined credible region, indicating low posterior predictive probability and thus low plausibility of being generated by the current model. Such deviations may signal data quality issues, unexpected patterns, or rare events that warrant further scrutiny.

\section{Simulation}

\begin{figure}
\centering
\caption{Simulation Scenarios. Left: observed trajectories for the first simulation data; Right: mean trajectory. In each figure, left is placebo and right is treatment.}
\label{fig:simScenario}
\includegraphics[width=0.48\textwidth]{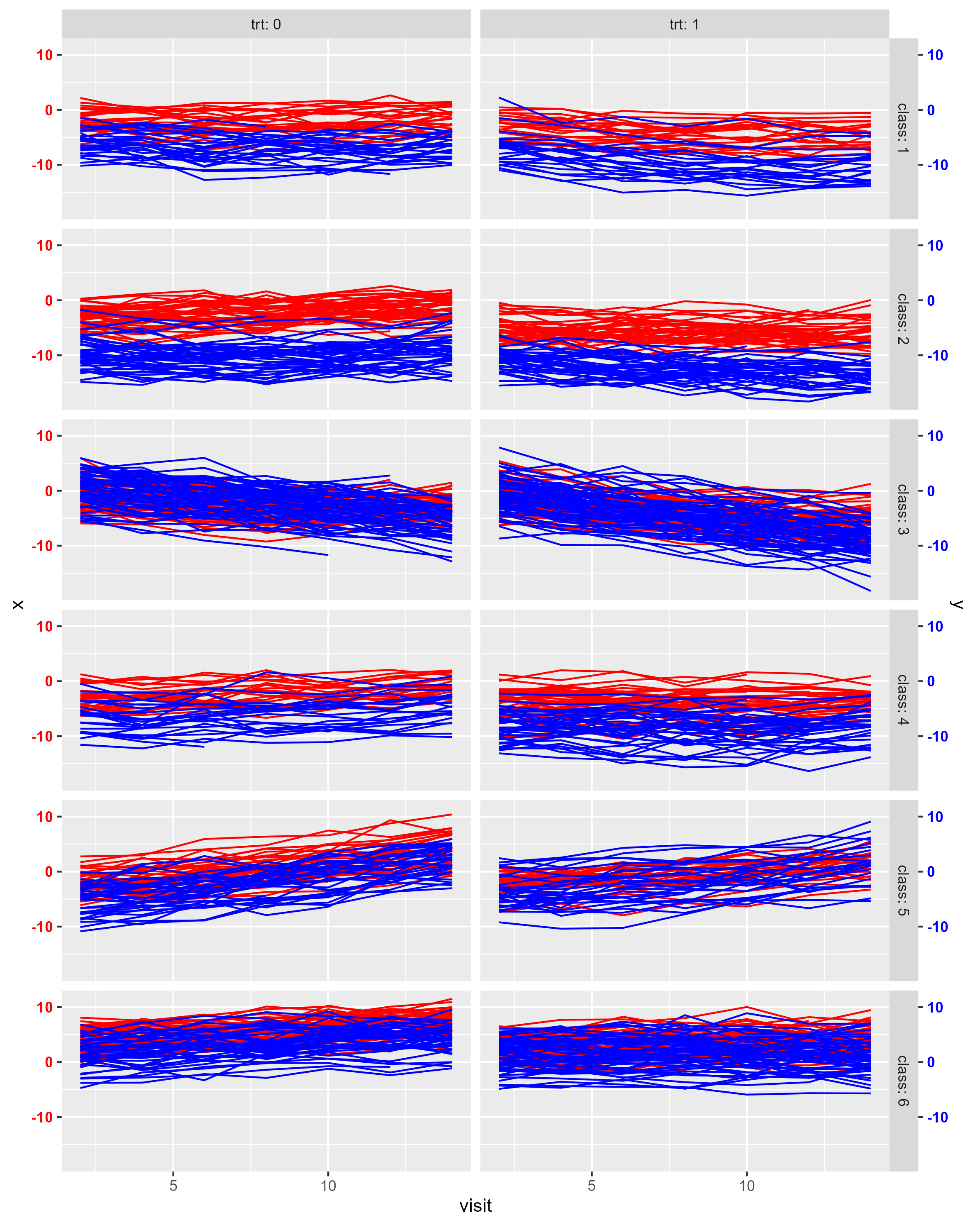}
\includegraphics[width=0.48\textwidth]{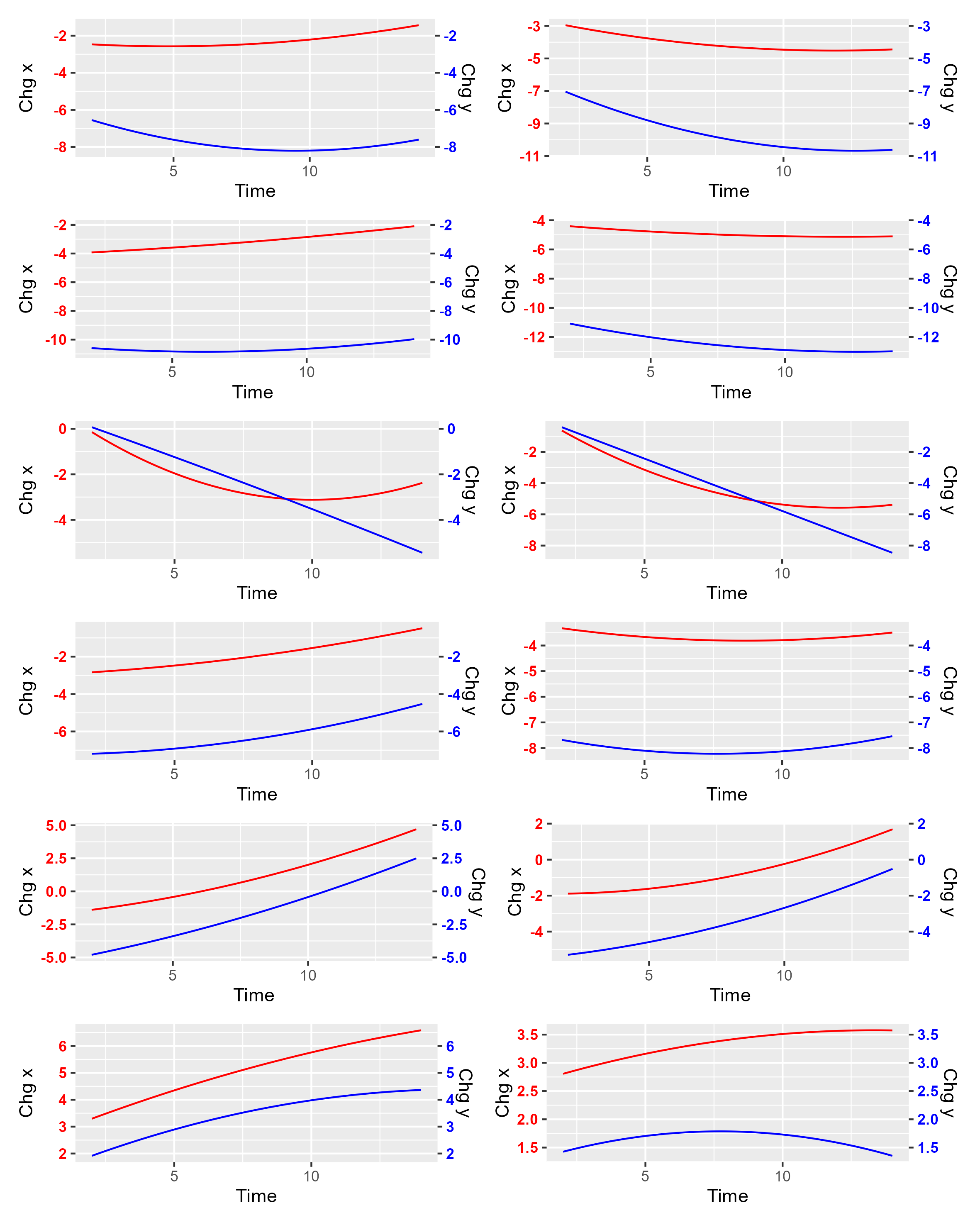}
\end{figure}

To simulate a pivotal study with a change from baseline endpoint, we consider a scenario with 50 sites and a total sample size of 700, split into training (70\%) and testing (30\%) datasets. Subjects are arbitrarily assigned to sites based on randomly set probabilities totaling to 1. There are six classes with true membership probabilities $(1/8, 1/8, 1/4, 1/4, 1/8, 1/8)$, each associated with site random effects (RE) categorized as low, low, moderate, moderate, high, high. Within each site, subjects are randomly assigned to treatment or placebo with a probability of 0.5. Visits occur at times $2, 4, 6, \ldots, 14$. Assume subjects have completed at least the first three site visits, including baseline, and the final visit is randomly selected from the remaining visits $(6, 8, \ldots, 14)$ with probabilities (0.01, 0.02, 0.04, 0.08, 0.85), representing an 85\% study completion rate. Six different scenarios are designed as listed in Table~\ref{tab:simScenario}, varying placebo effects from treatment (i.e. adding $0.25$ to $\beta_{1c}^x$ and $\beta_{1c}^y$ and adding $-0.0025$ to $\beta_{2c}^x$ and $\beta_{2c}^y$ ), resulting in a total of 12 distinct classes as shown in Figure~\ref{fig:simScenario}. The simulated trajectories are generated based on specified model settings in (\ref{eq:modelSetup1}) and (\ref{eq:modelSetup2}).

With a total of 100 replicates, $C=30$ clusters are set for training, and posterior prediction is conducted for subjects with at least three post-baseline visits in testing. The posterior prediction of the third post-baseline observation, conditional on the first two, is summarized in Table ~\ref{tab:simResults}. Due to high computational burden, each replicate involves a burn-in of 50,000 iterations, with an additional 1,000 iterations used for posterior inference. Grid regions to derive the 80\% credible region are intentionally reduced for computational ease and are based on $\{\text{(}-16,-14\text{]},\text{(}-14,-12\text{]},\cdots, \text{(}14,16\text{]}\}$  for $x$ and $\{\text{(}-24,-22\text{]}, \text{(}-22,-20\text{]},\cdots, \text{(}14,16\text{]}\}$ for $y$. Assessment includes bias, mean squared error, and the proportion of subjects in the posterior predictive region using two algorithms, both of which perform well in recovering the true values.

\begin{table}[]
    \centering
    \begin{tabular}{c|c|c}
    \hline
         & Algorithm 1: Branching out & Algorithm 2: HDR \\ \hline
     Proportion of in the regions: mean (SD) &  $85.74\% (3.25\%)$ & $85.66\% (3.25\%)$ \\  \hline
     Bias: mean (SD) & $1.78\% (0.200\%)$ & $1.81\% (0.200\%)$\\ \hline
     sqrt. MSE: mean (SD) & $2.27\% (0.216\%)$ & $2.28\% (0.214\%)$ \\ \hline
    \end{tabular}
    \caption{Simulation results from $100$ replications}
    \label{tab:simResults}
\end{table}

\begin{table}[]
\centering
\caption{Simulation Scheme}
\label{tab:simScenario}
{\small
\begin{tabular}{c|c|c|c|c|c|c|c}
\hline
 Parameter & Class 1 & Class 2 & Class 3 & Class 4 & Class 5 & Class 6 & Common \\ \hline  
 $\pi_{ic}$ & 1/8 & 1/8 & 1/4 & 1/4 & 1/8& 1/8 &  \\ \hline
 site RE & low & low & moderate & moderate & high & high &  \\ \hline
 $\tau_{sc}^x$ & 1/0.75& 1/0.75 & 1/1.25 & 1/1.25 &1/2 &1/2 &\\ \hline
 $\beta_{0,\text{base}}^x$ & & & & & & & -0.4 \\ \hline
 $\mu_{0,\text{base}}^x$ & & & & & & & 10 \\ \hline
 $\tau_{0,\text{base}}^x$ & & & & & & & 1/12 \\ \hline
  $\tau_{w}^x$ & & & & & & & 1 \\ \hline
   $\tau_{e}^x$ & & & & & & & 1 \\ \hline
 $\beta_{0c}^x$ &1.74& -0.10&  5.53&  1.01&  2.16&  6.51 &\\ \hline
 $\beta_{1c}^x$ &-0.38& -0.17& -1.18& -0.19& -0.07&  0.16 &\\ \hline
 $\beta_{2c}^x$ & 0.016&  0.007&  0.049&  0.011&  0.023& -0.006 &\\ \hline
$\beta_{0,\text{base}}^y$ & & & & & & & -0.6\\ \hline
$\mu_{0,\text{base}}^y$ & & & & & & & 16 \\ \hline
 $\tau_{0,\text{base}}^y$ & & & & & & & 1/12 \\ \hline
   $\tau_{w}^y$ & & & & & & & 1.25 \\ \hline
   $\tau_{e}^y$ & & & & & & & 1.25 \\ \hline
   $\tau_{sc}^y$ & 1/0.9375& 1/0.9375 &1/1.5625 & 1/1.5625 & 1/2.5 & 1/2.5&\\ \hline
 $\beta_{0c}^y$ &4.05& -0.69& 10.52&  2.37&  4.02& 10.73 &\\ \hline
 $\beta_{1c}^y$ &-0.81& -0.43& -0.67& -0.26&  0.11&  0.17 &\\ \hline
 $\beta_{2c}^y$ & 0.032&  0.017&  0.000&  0.017&  0.018& -0.011 &\\ \hline
\end{tabular}
}
\end{table}

\section{Data Application}

Atopic dermatitis, a prevalent condition often originating in early childhood, is commonly assessed for efficacy using endpoints such as the Eczema Area and Severity Index (EASI) score, Investigator’s Global Assessment (IGA) score, and Scoring Atopic Dermatitis (SCORAD) index. Both EASI and SCORAD are continuous endpoints, with scores ranging from 0 to 72 and 0 to 103, respectively, where higher scores indicate greater disease severity. Conversely, IGA is a discrete endpoint, graded on a five-point scale ranging from 0 (clear) to 4 (severe AD).

Subsequent analyses focused on EASI versus IGA, and EASI versus SCORAD, utilizing data from a Phase 3 clinical trial, which was divided into training (70\%) and testing (30\%) sets. Results obtained from both algorithms, along with visualizations based on the best cluster configuration determined through posterior samplings using the least squares method \citep{dahl2006model}, were presented.

\subsection{EASI and IGA}

For training, the data comprised changes from baseline in both EASI and IGA scores. We utilized 30 latent classes ($C=30$). The MCMC sampler was run for 50,000 iterations, with the final 2,000 iterations used for posterior inference. The test dataset included 243 subjects (30\% of the total), and we assessed two scenarios.

\begin{figure}
\centering
\includegraphics[width=0.9\textwidth]{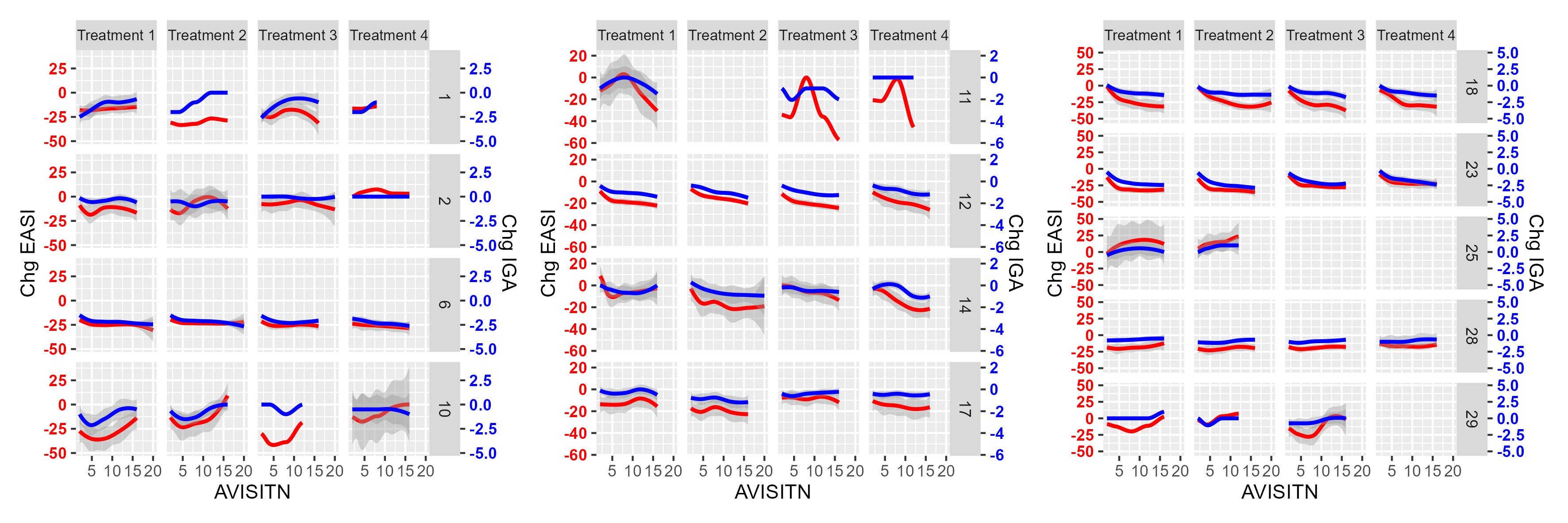}
\includegraphics[width=0.9\textwidth]{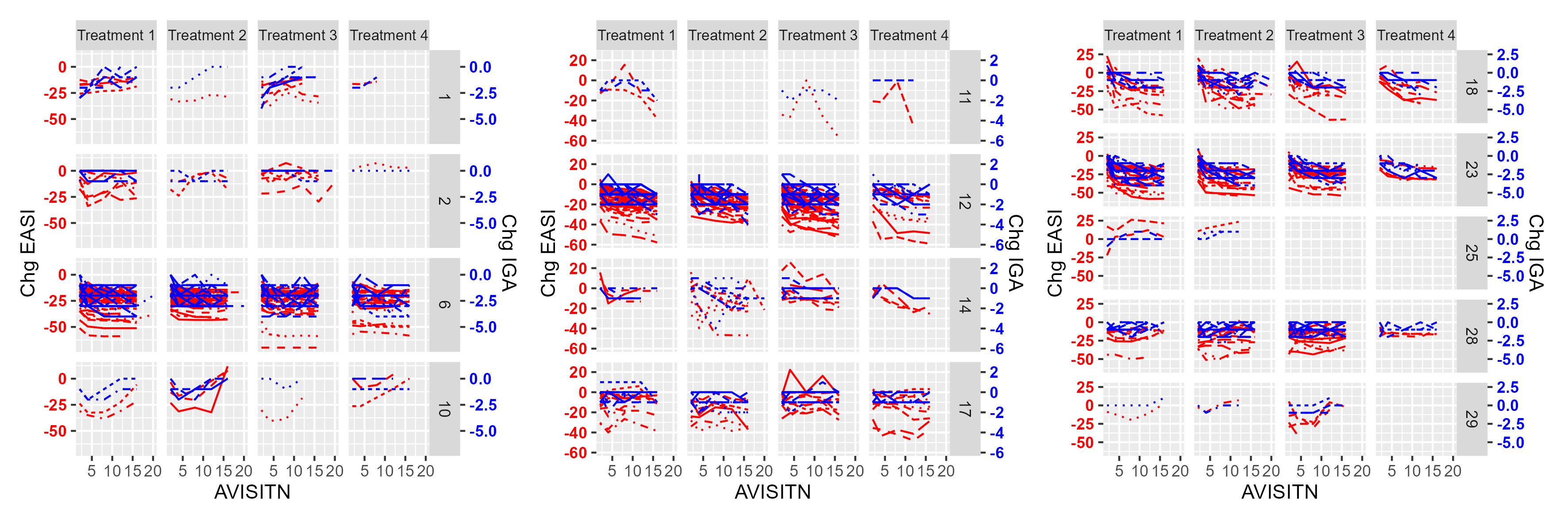}
\caption{EASI and IGA: Classification (number of subjects) of $570$ subjects from best configuration \citep{dahl2006model}: 1(12), 2(15), 6(157), 10(8), 11(4), 12(146), 14(19), 17(38), 18(39), 23(77), 25(4), 28(45), 29(6). Top: Mean Trajectory; Bottom: Individual Trajectory. Note that for the bottom of individual trajectories,  it is faceted by treatment for easier visualization, and training was based on blinded data.}
\label{fig:fitclass}
\end{figure}

Scenario 1 involved predicting outcomes at Week 2 based on baseline values of EASI and IGA, with 240 subjects having data available at Week 2. Scenario 2 aimed to predict outcomes at the fourth visit based on observed EASI and IGA values from the first three post-baseline visits, including 228 subjects with data for more than four postbaseline visits. The classification using the best iteration of MCMC is shown in Figure~\ref{fig:fitclass} and the results on the test data are summarized in Table~{\ref{tab:app1pp}}.

\begin{figure}[]
    \includegraphics[width=0.5\textwidth]{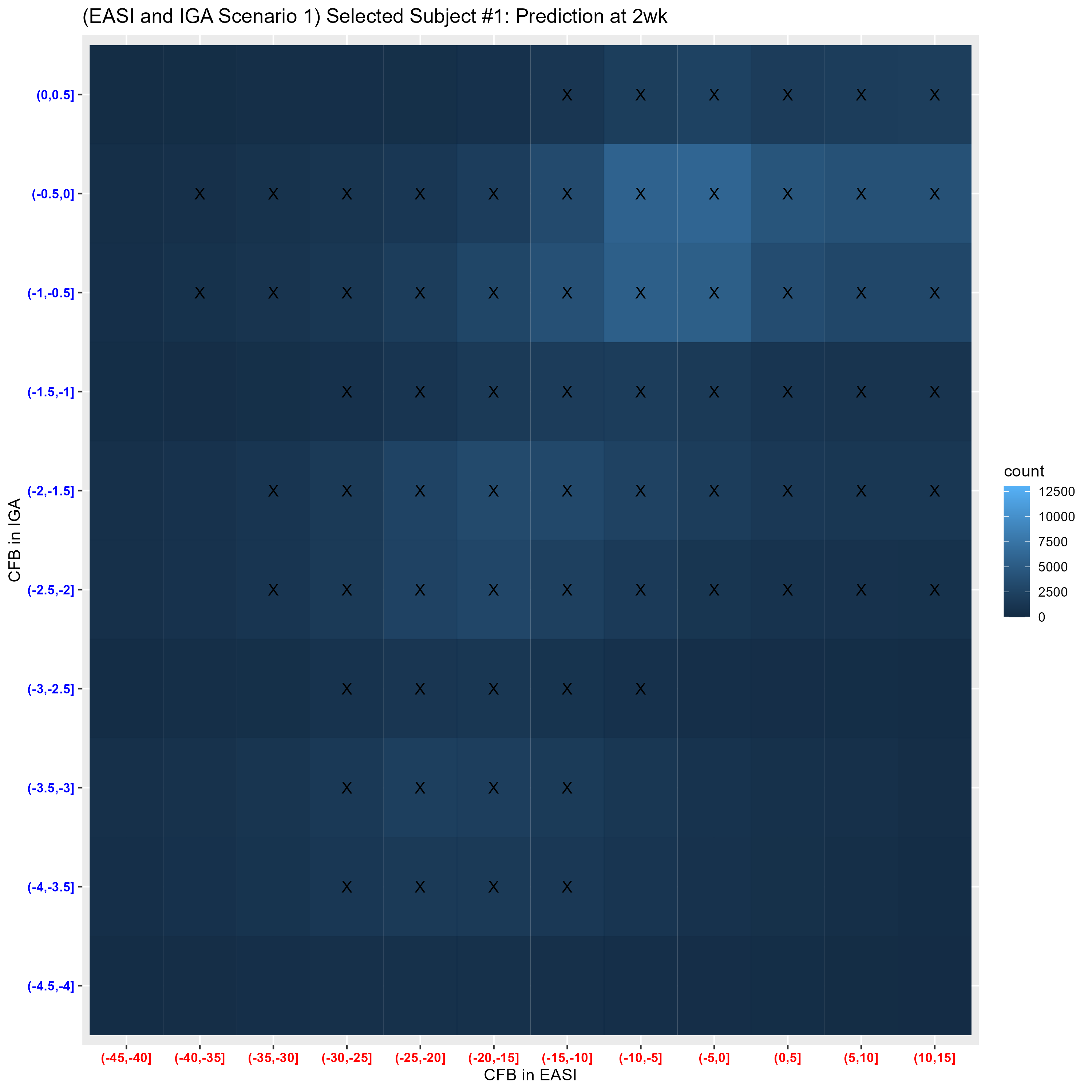}
    \includegraphics[width=0.5\textwidth]{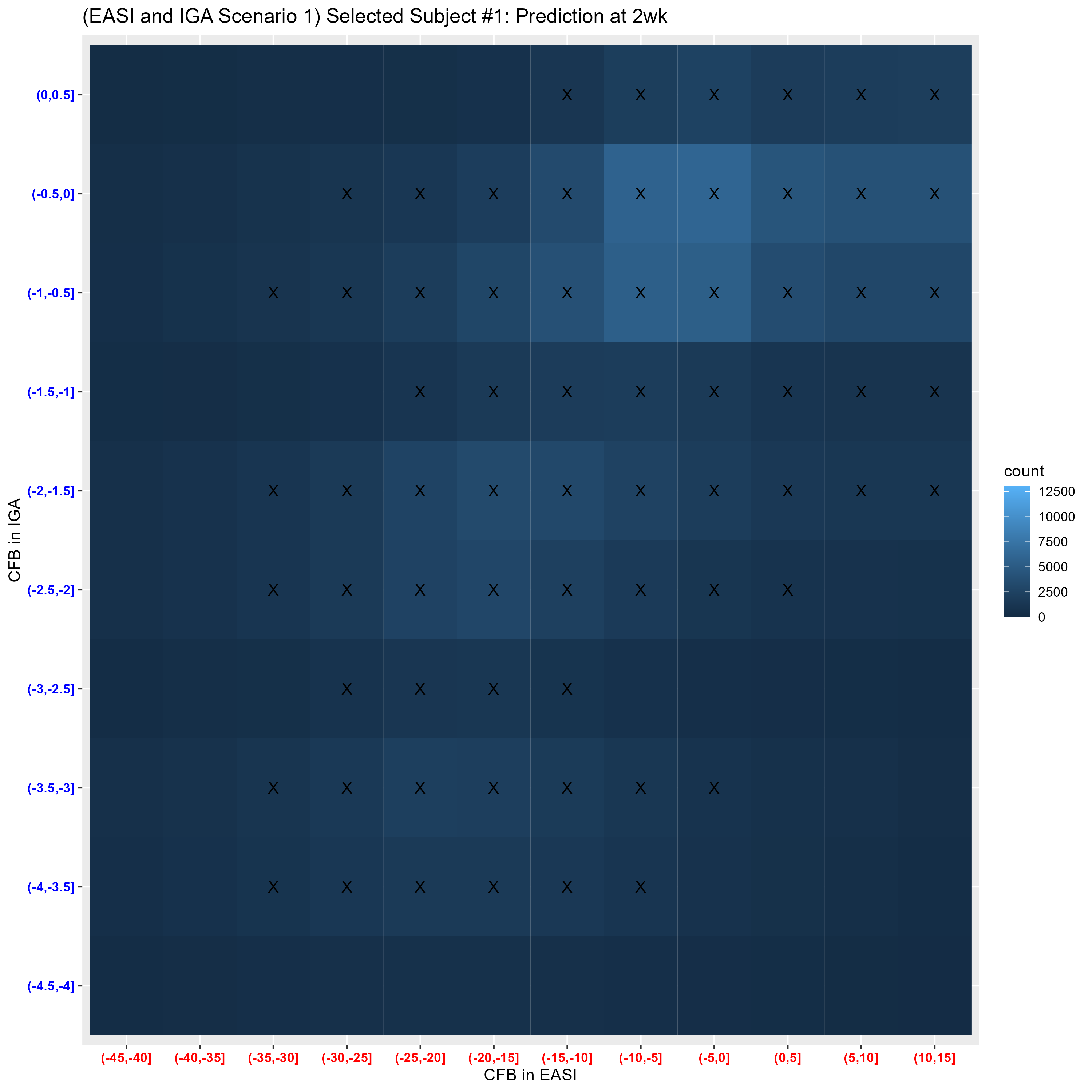}
    \caption{Prediction Scenario 1 (EASI and IGA): $80\%$ credible regions for Scenario 1 Selected Subject $\#1$ at Week 2 conditional on baseline. The $80\%$ credible regions are identified with X over the regions. Left: Algorithm 1 (Branching out); Right: Algorithm 2: HDR. Baseline (EASI:21, IGA:3); True value at Week 2(EASI:-10, IGA:0). The observation at Week 2 is contained in the credible regions for both algorithms.} 
    \label{fig:app1pp1subj1}
\end{figure}

\begin{figure}[!htp]
    \includegraphics[width=0.5\textwidth]{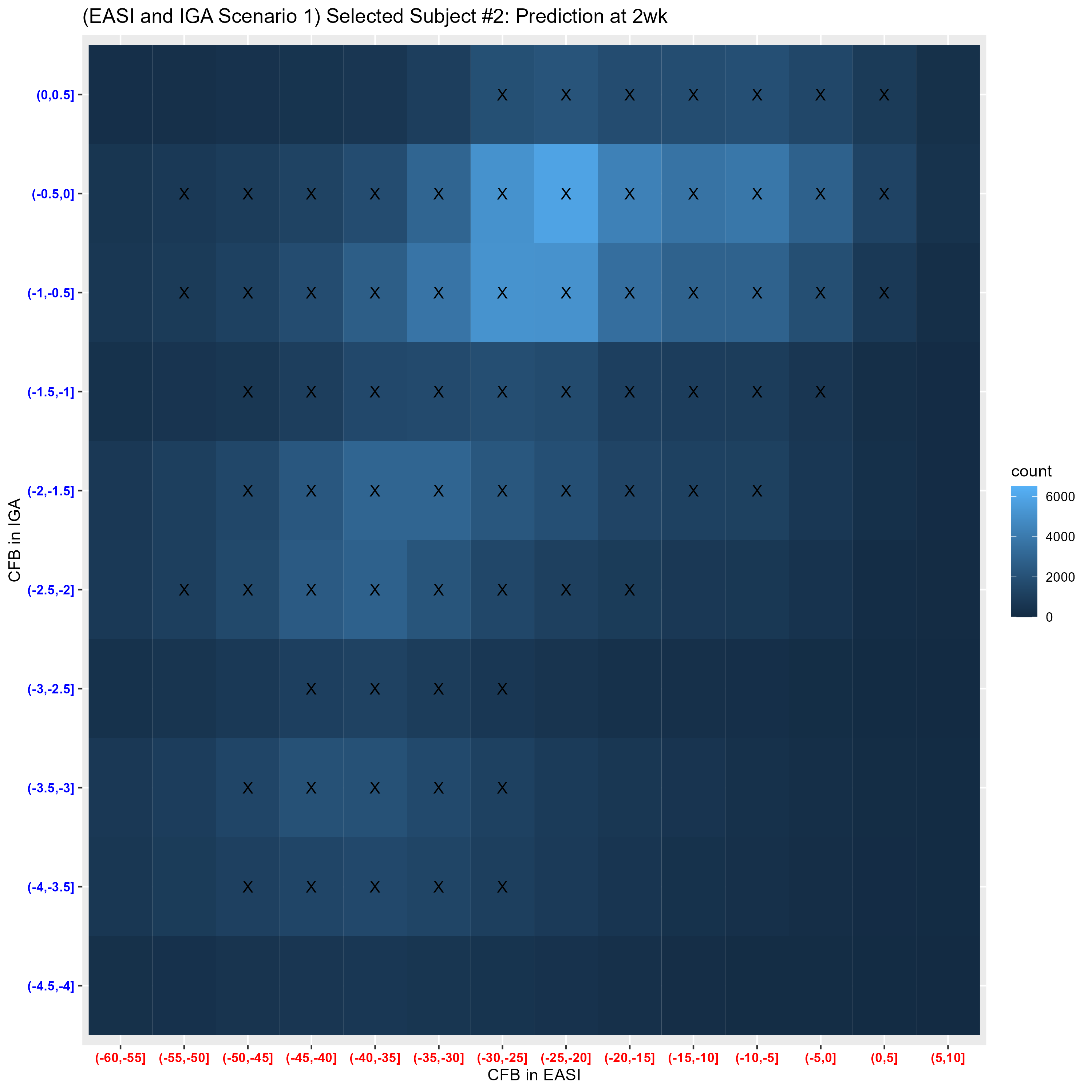}
 \includegraphics[width=0.5\textwidth]{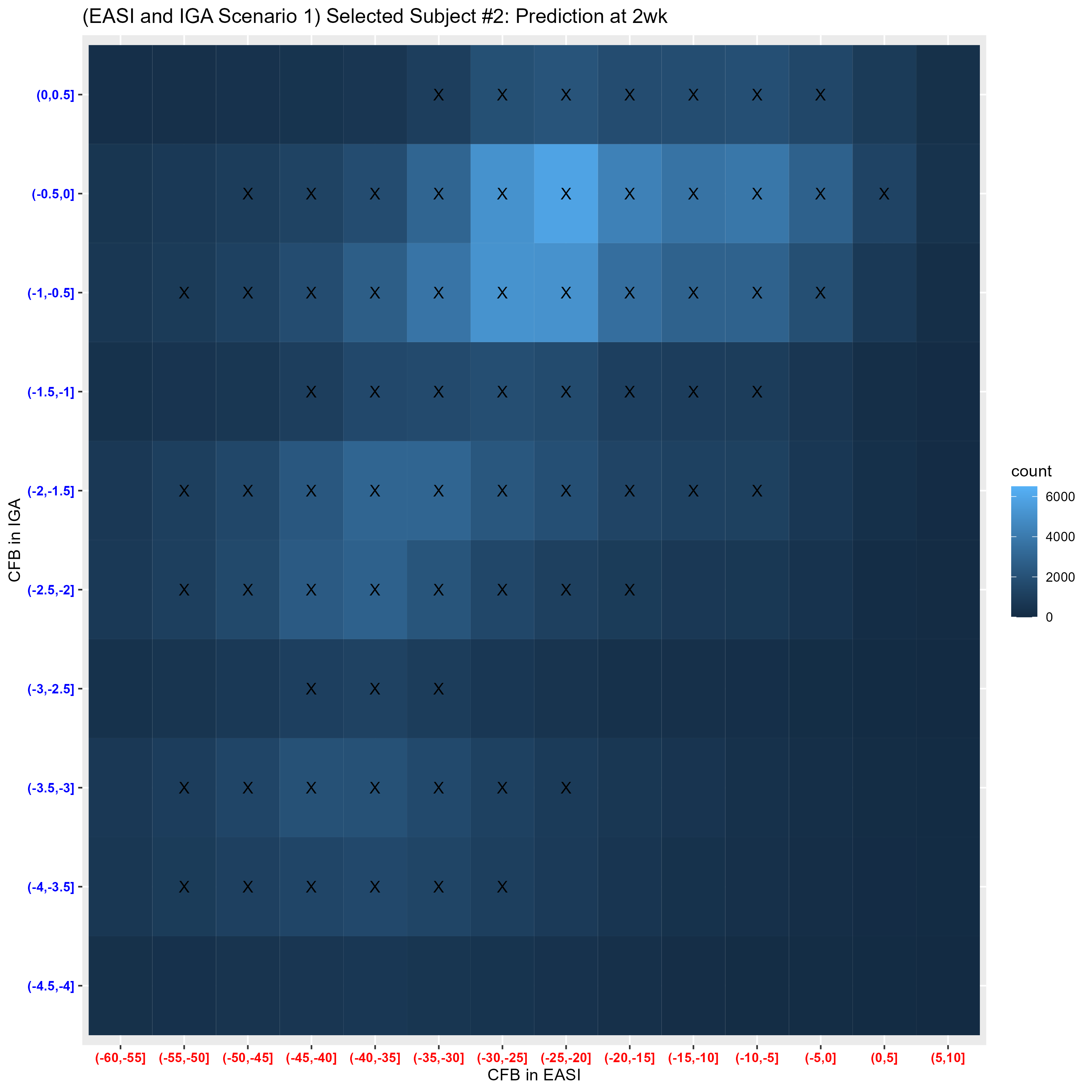}
    \caption{Prediction Scenario 1  (EASI and IGA): $80\%$ credible regions for Scenario 1 Selected Subject $\#2$ at Week 2 conditional on baseline. The $80\%$ credible regions are identified with X over the regions. Left: Algorithm 1 (Branching out); Right: Algorithm 2: HDR. Baseline (EASI:43.2, IGA:3); True value at Week 2 (EASI:-15.3, IGA:-1). The observation at Week 2 is contained in the credible regions for both algorithms. }
     \label{fig:app1pp1subj2}
\end{figure}

\begin{figure}[!htp]
    \includegraphics[width=0.5\textwidth]{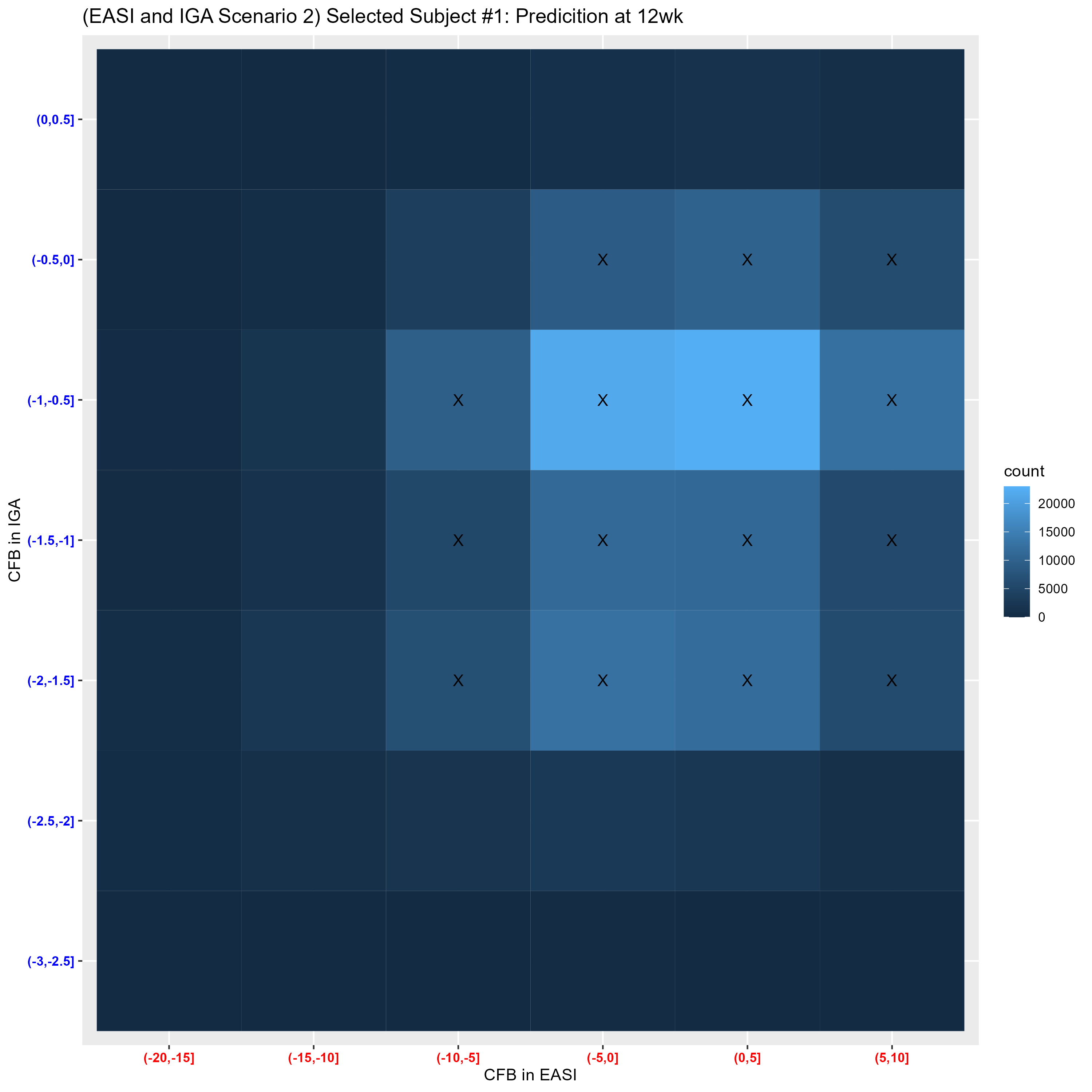}
    \includegraphics[width=0.5\textwidth]{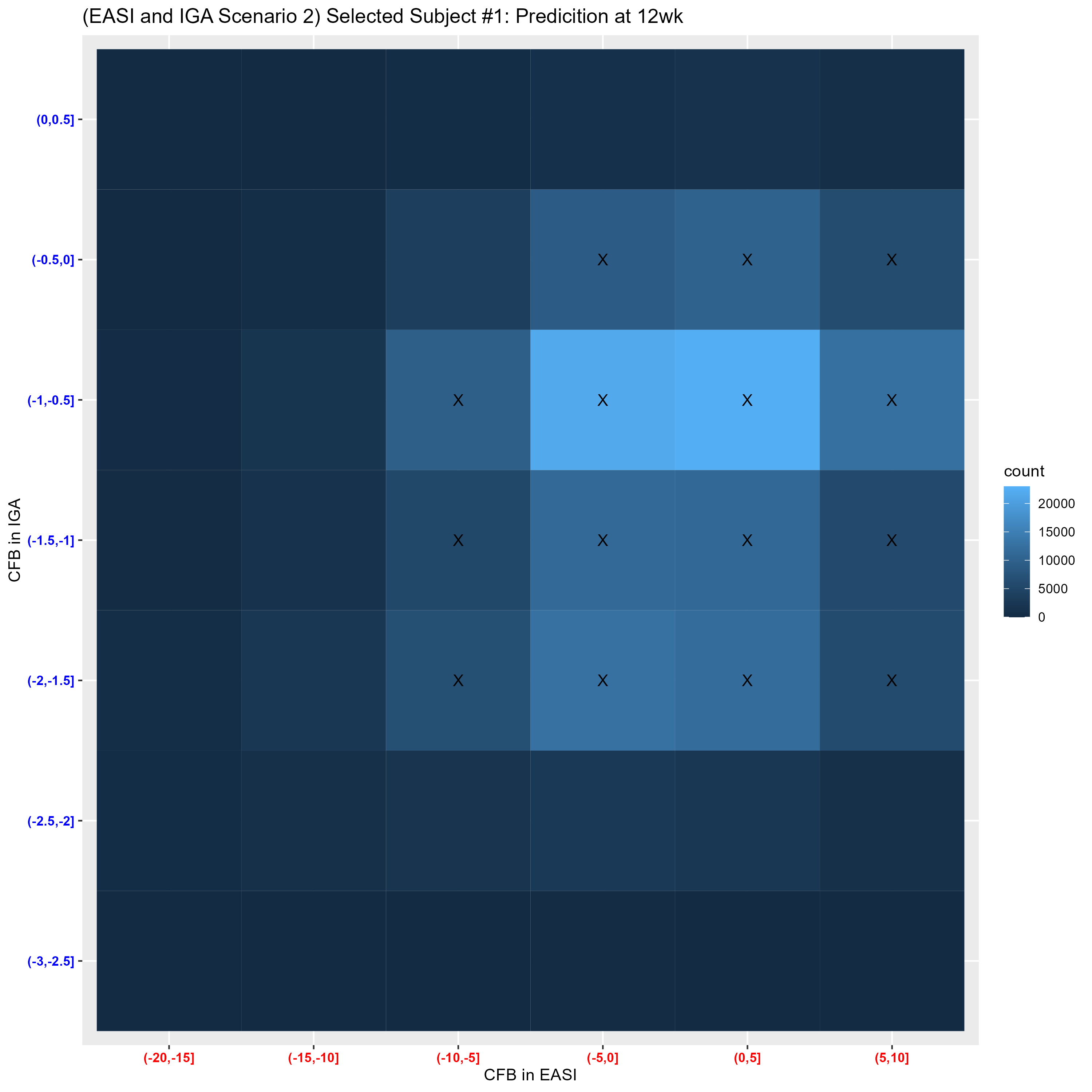}
    \caption{Prediction Scenario 2  (EASI and IGA): $80\%$ credible regions for Scenario 2 Selected Subject $\#1$ at Week 12 conditional on Weeks 2, 4, and 8. The $80\%$ credible regions are identified with X over the regions. Left: Algorithm 1 (Branching out); Right: Algorithm 2: HDR. Baseline (EASI:17.6, IGA:4); True value  at Week 12 (EASI:-14, IGA:-1). The observation at Week 12 is not contained in the credible regions for both algorithms.}
     \label{fig:app1pp2subj1}
\end{figure}

\begin{figure}[!htp]
    \includegraphics[width=0.5\textwidth]{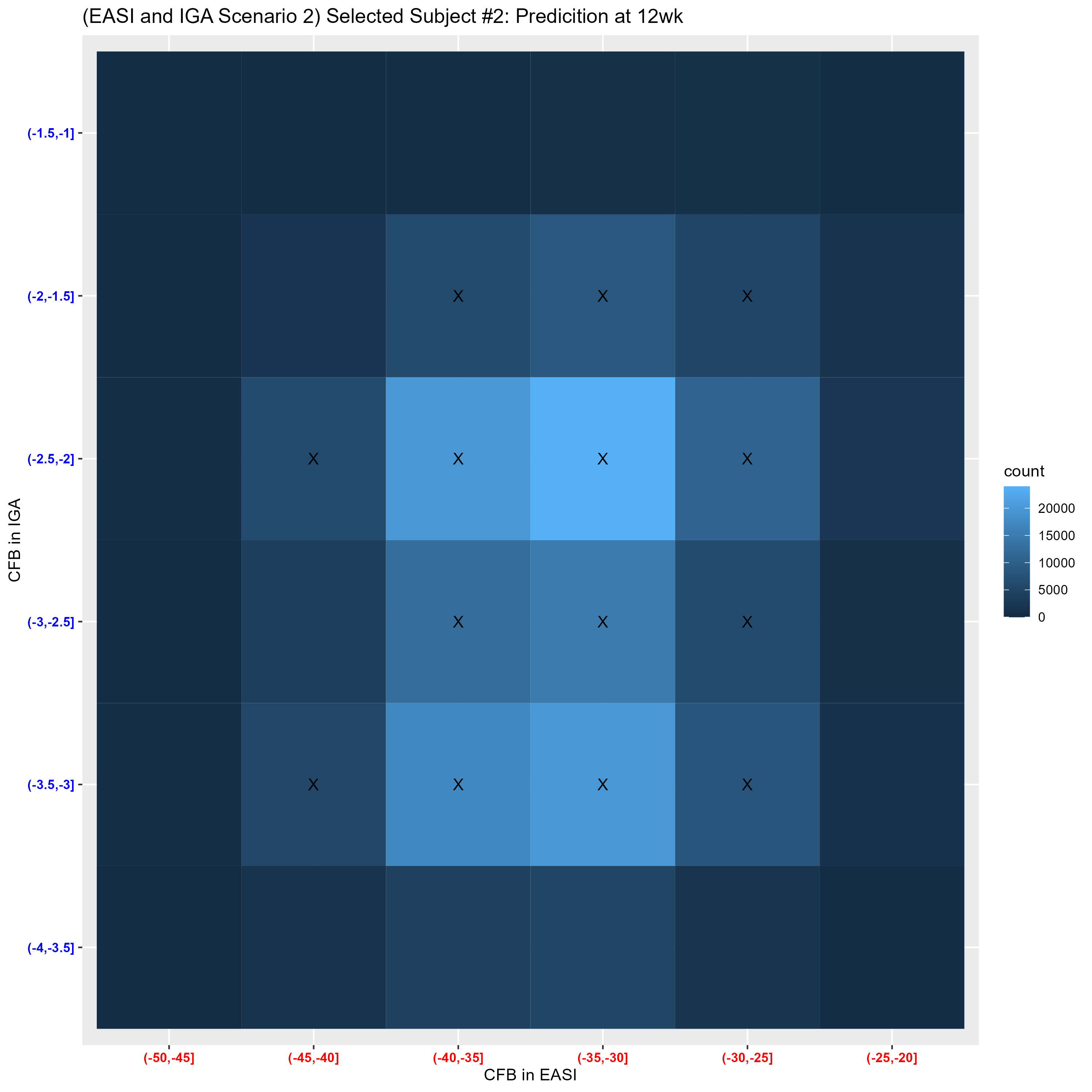}
    \includegraphics[width=0.5\textwidth]{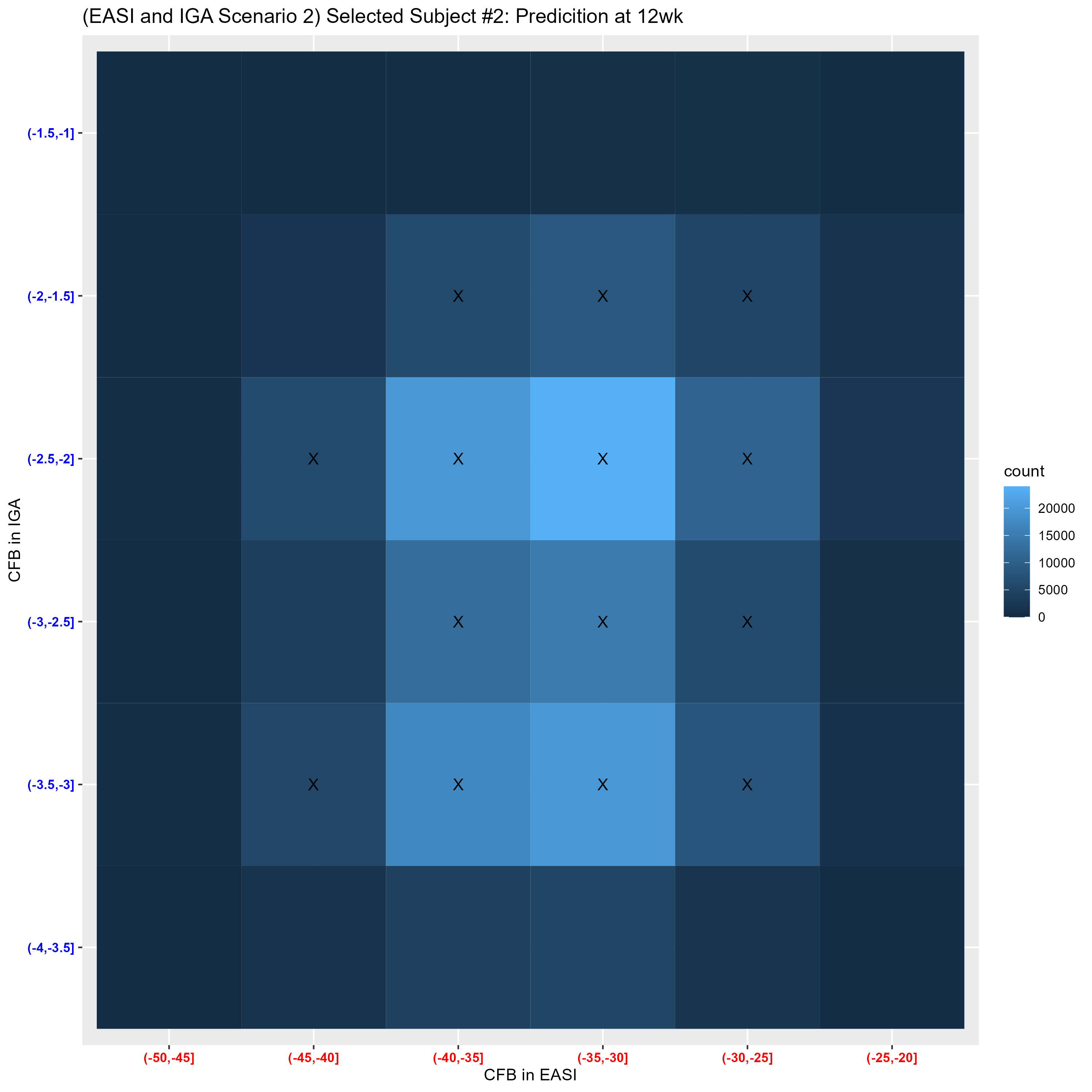}
    \caption{Prediction Scenario 2  (EASI and IGA): $80\%$ credible regions for Scenario 2 Selected Subject $\#2$ at Week 12 conditional on Weeks 2, 4, and 8. The $80\%$ credible regions are identified with X over the regions. Left: Algorithm 1 (Branching out); Right: Algorithm 2: HDR. Baseline (EASI:34.8, IGA:4); True value at Week 12 (EASI:-33.8, IGA:-3). The observation at Week 12 is contained in the credible regions for both algorithms.}
     \label{fig:app1pp2subj2}
\end{figure}

For each scenario, two subjects were selected to visualize the regions derived using both region algorithms at the 80\% credible level (see Figures ~\ref{fig:app1pp1subj1},\ref{fig:app1pp1subj2},\ref{fig:app1pp2subj1},\ref{fig:app1pp2subj2}). When conditioning on baseline values, the proportion of true values within the posterior predictive region was notably higher compared to predictions for the fourth post-baseline visit based on the first three observations. Both algorithms showed comparable performance, with subtle differences observed in the selected regions (see Figures ~\ref{fig:app1pp1subj1}, \ref{fig:app1pp1subj2}). When fewer high-density regions were involved, the two algorithms produced similar results (see Figures \ref{fig:app1pp2subj1},\ref{fig:app1pp2subj2}).

\begin{table}[]
\caption{EASI and IGA: Posterior Prediction Results.}
\label{tab:app1pp}
\small
\begin{tabular}{c|c|c|c|c}
\hline\hline
   & Algorithm & prob. of in credible region & bias of credible level & square root of MSE \\ \hline
Scenario 1& 1: branching out & 0.954 & 0.0014 & 0.0017 \\ \hline
   & 2: HDR & 0.971 & 0.0024 & 0.0028 \\ \hline
Scenario 2 & 1:branching out  & 0.772 & 0.0088 & 0.0106 \\ \hline
   & 2:HDR & 0.785 & 0.0093 & 0.0110 \\ \hline\hline
\end{tabular}
\end{table}

\subsection{EASI and SCORAD}

\begin{figure}
\includegraphics[width=0.95\textwidth]{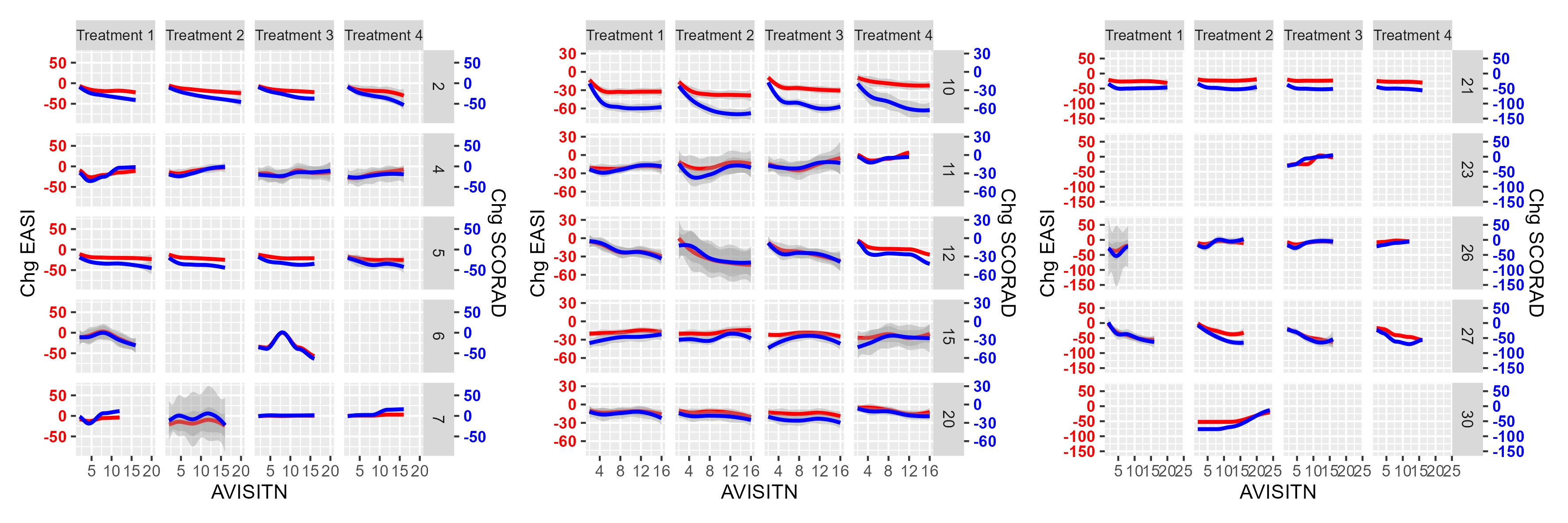}
\includegraphics[width=0.95\textwidth]{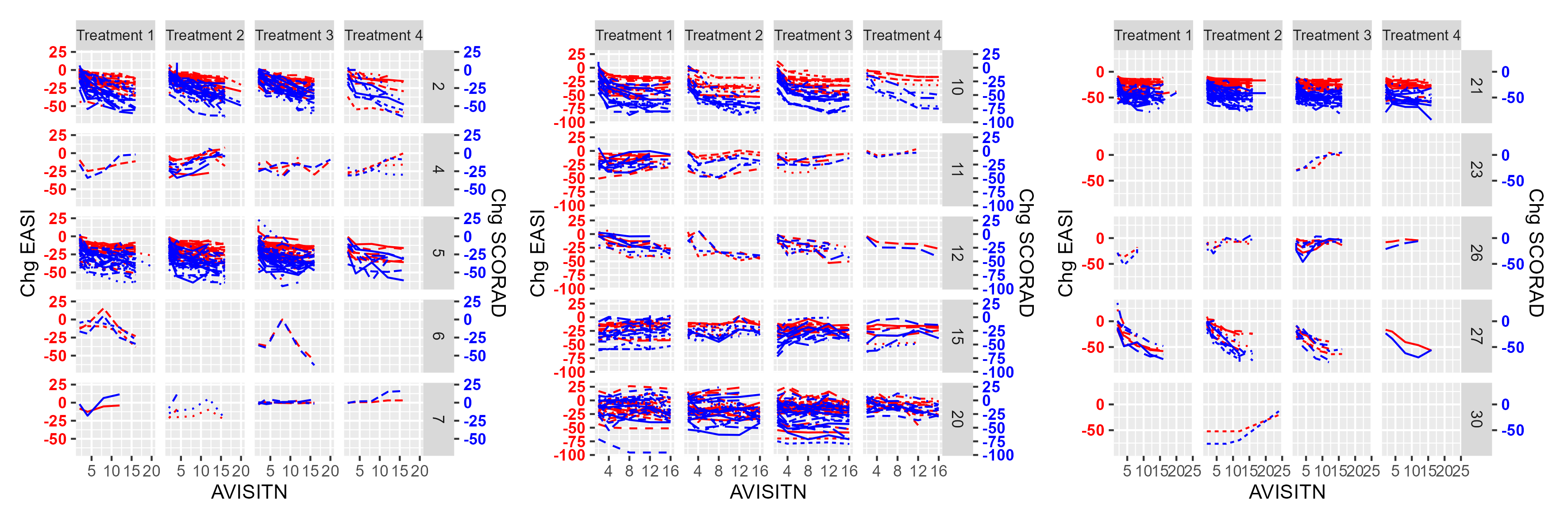}
 \caption{EASI and SCORAD: Classification (number of subjects) of $566$ subjects from the best configuration \citep{dahl2006model}: 2(84), 4(10), 5(100), 6(3), 7(7), 10(52), 11(20), 12(14), 15(41), 20(80), 21(123), 23(1), 26(9), 27(21), 30(1). Top: mean trajectory; Bottom: individual trajectory. Note that for the bottom of individual trajectories,  it is faceted by treatment for easier visualization, and training was based on blinded data.}
 \label{fig:fitclass2}
\end{figure}

The data used for fitting included changes from baseline in both EASI and SCORAD scores. The dataset was divided into training (70\%) and testing (30\%) sets. For training, the number of latent classes was set to $C=30$. The total sampler for MCMC was 50,000 iterations, with the last 2,000 used for posterior inference. The classification using the best iteration of MCMC is shown in Figure~\ref{fig:fitclass2}.

For the 243 subjects in the test data (30\%), we assessed two scenarios:
\begin{itemize}
    \item {\bf Scenario 1}: Prediction at Week 2, conditional on baseline values of EASI and SCORAD. This included 236 subjects with data available at Week 2.
    \item {\bf Scenario 2}: Prediction at the fourth visit, conditional on observed values of EASI and SCORAD at the first three post-baseline visits. This included 226 subjects with data for more than four visits.
\end{itemize}

\begin{figure}[!htp]
    \includegraphics[width=0.5\textwidth]{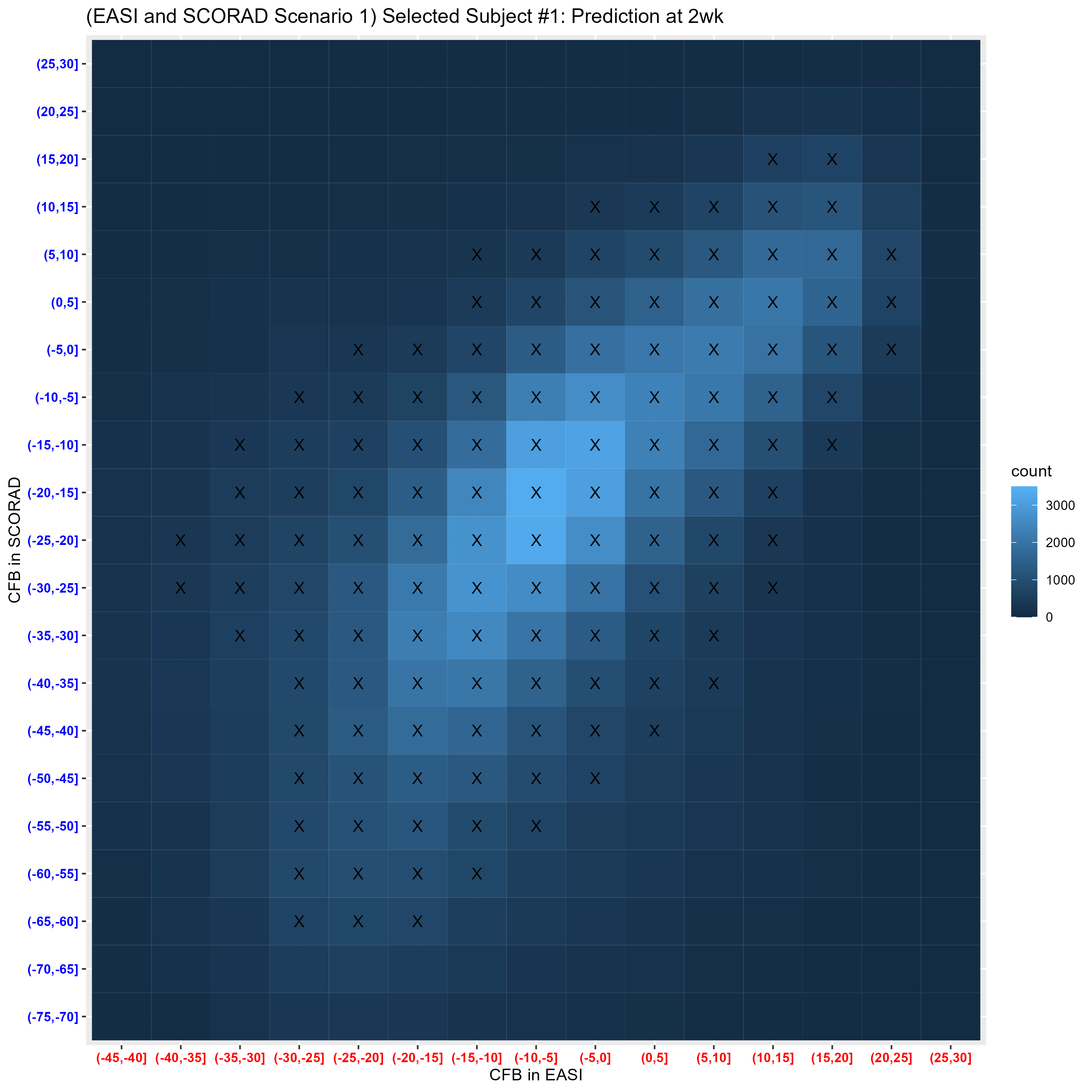}
    \includegraphics[width=0.5\textwidth]{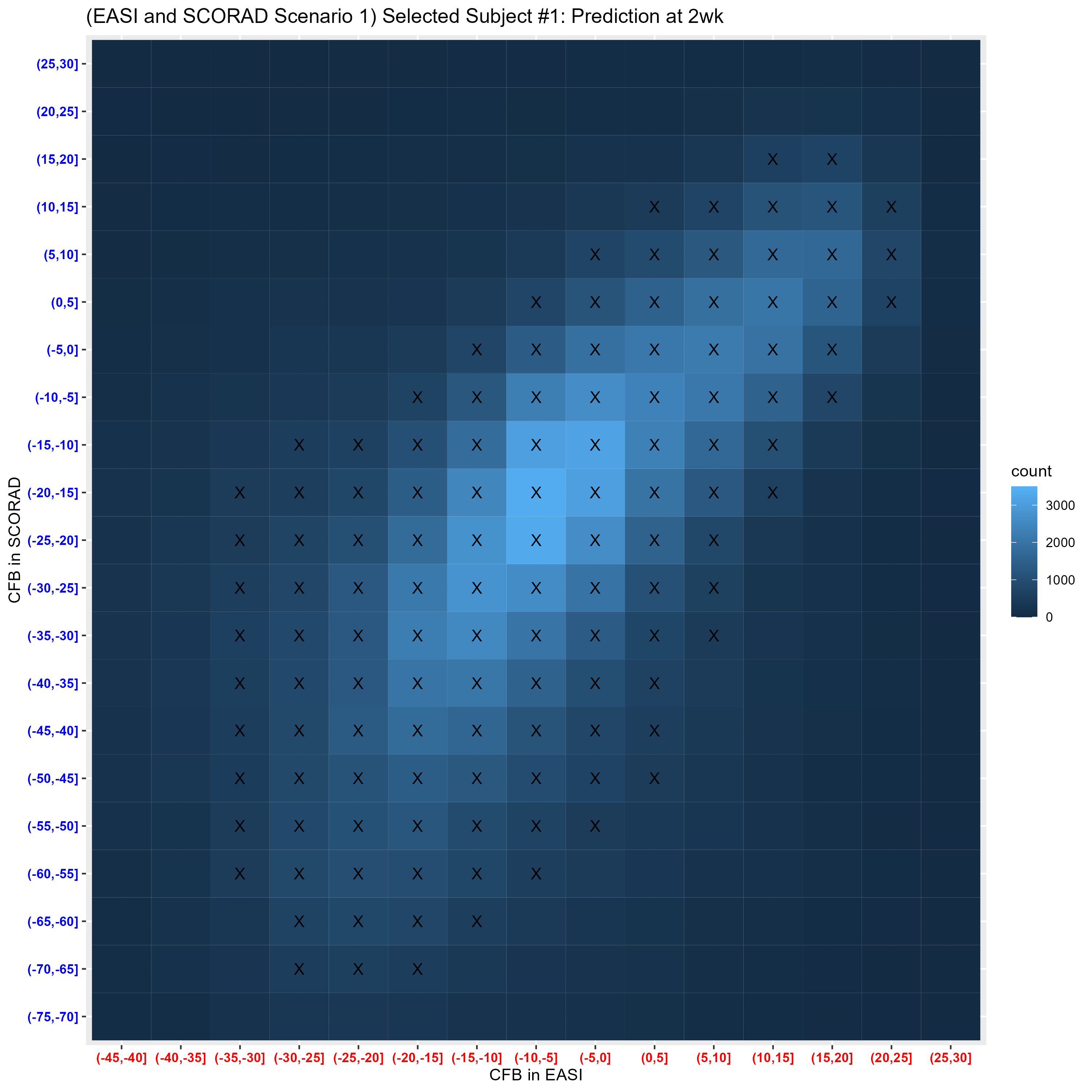}
    \caption{Prediction Scenario 1 (EASI and SCORAD): $80\%$ credible regions for Scenario 1 Selected Subject $\#1$ at Week 2.  Left: Algorithm 1 (Branching out); Right: Algorithm 2: HDR. Baseline (EASI:17.6, SCORAD:54.7); True value at Week 2 (EASI:0.4, SCORAD:-11.6). The observation at Week 2 is contained in the credible regions for both algorithms.}
    \label{fig:app2pp1subj1}
\end{figure}

The results are summarized in Table~{\ref{tab:app2pp}}. For each scenario, two subjects were selected for visualization of selected regions (see Figures ~\ref{fig:app2pp1subj1}, \ref{fig:app2pp1subj2},\ref{fig:app2pp2subj1},\ref{fig:app2pp2subj2},\ref{fig:app2pp2subj3} in the appendix) using both region algorithms to derive the 80\% credible level.

The observations are consistent with the previous analysis. When conditioning on the baseline value, the proportion of true values within the posterior predictive region is very high compared to the prediction of the fourth post-baseline visit, conditioned on the three previous observations. The two algorithms are comparable, with subtle differences in the selected regions (see Figures ~\ref{fig:app2pp1subj1}, \ref{fig:app2pp1subj2} in the appendix). With fewer high-density regions, both algorithms yield the same results (see Figures \ref{fig:app2pp2subj1},\ref{fig:app2pp2subj2},\ref{fig:app2pp2subj3} in the appendix). 

Additionally, compared to the previous analysis of EASI and IGA, the probability of recovering values within the credible region is higher for Scenario 2. This is expected since SCORAD can be better fitted using our quadratic model than IGA, which is more on a discrete scale.

\begin{table}[]
\caption{EASI and SCORAD: Posterior Prediction Results}
\label{tab:app2pp}
\centering
\small
\begin{tabular}{c|c|c|c|c}
\hline\hline
   & Algorithm & prob. of in credible region & bias of credible level & square root of MSE \\ \hline
Scenario 1& 1: branching out & 0.949 & 0.00072 & 0.00085 \\ \hline
   & 2: HDR & 0.945 & 0.0015 & 0.0019 \\ \hline
Scenario 2 & 1:branching out  & 0.876 & 0.0053 & 0.0067 \\ \hline
   & 2:HDR & 0.885 & 0.0061 & 0.0073 \\ \hline\hline
\end{tabular}
\end{table}

\section{Discussion}

In this research, we presented a tool for identifying anomalous data through a model-based approach that features both primary and key secondary endpoints. By constructing a joint model of multiple endpoints through risk-modified latent classes, we can learn distinct subgroup behaviors. Individualized posterior predictions can be made by marginalizing the latent classes and random effects. However, this method is currently limited to key efficacy endpoints and focuses on different trends within those instruments. Data monitoring is a collective process that should incorporate multiple tools, each intended for different objectives.

Our framework is flexible and can be easily extended. The proposed setup primarily considers continuous data. For binary data, it can be adapted to a generalized linear mixed model. Additionally, our method can be extended to handle multiple endpoints. In Section 4, we separately analyzed EASI and IGA, and EASI and SCORAD. It is also possible to jointly analyze all three endpoints in one run. Furthermore, our setup currently considers quadratic curvature, but it can be extended to incorporate fractional polynomials as the core model to accommodate a wide variety of trajectories. For example, 
\begin{equation*}
    \beta_{0} +\beta_{1}t^{p_1} +\beta_2t^{p_2}
\end{equation*}
or under $p_1 = p_2$
\begin{equation*}
    \beta_{0} +\beta_{1}t^{p_1} +\beta_2t^{p_1}\ln t,
\end{equation*}
where $p_1$ and $p_2$ can be selected from $-2,-1,-0.5,0,0.5,1,2,3$. For model selection, existing methods like DIC (Deviance Information Criterion) and LPML (Log Pseudo Marginal Likelihood) can be utilized.

While our proposal offers significant functionality, such as individualized prediction for detecting anomalous data, it does have certain limitations. The model incorporates risk factors, such as site, to modify latent class assignments. However, a clear method for quantitatively assessing overall site risk factors and systematically rating suspicious sites is lacking. Additionally, there is no robust mechanism to impose stronger controls on flagging datasets from these sites, even when they do not appear anomalous. Further assessment of identified anomalous data can be conducted by examining the frequency and percentage of ``anomalous data'' from each site, but this approach remains qualitative rather than automated and quantitative.

At the onset of a study, accumulating sufficient data can be challenging, which may render predictions using the proposed method potentially unstable. Therefore, it is advisable to use other tools in conjunction with the proposed method during the early stages of the study. Effective monitoring often requires the simultaneous implementation of multiple tools. Initially, historical data from other studies can be used to inform predictions and enhance precision. However, as more data become available within the current study, reliance on historical data should decrease to avoid conflicts with site-specific influences.

Frequent monitoring using other practices may be necessary at the start of the study, gradually becoming less intensive over time. Additionally, the probability threshold for credible intervals can be relaxed eventually. 

\section*{Software}
The code can be requested at yuxi.zhao@pfizer.com or found at Github \url{https://github.com/helenjad/CTAnomalyScreen.git}.

\section*{Declaration of conflicting interests}
The author(s) declared no conflict of interests.

\section*{Funding}
The author(s) received no financial support for the research, authorship, and/or publication of this article.

\bibliographystyle{agsm}
\bibliography{reference}

\appendix 

\section{Priors and Hyperpriors}
As specified, the model is shown as:
\begin{align*}
x_{i,j} (t_{i,j}^x)\mid \{z_{i}=c\} =& \beta_{0c}^x  +  \beta_{0,base}^x Baseline^x  +
\beta_{1c}^x t_{i,j}^x + \beta_{2c}^x (t_{i,j}^x)^2 +  v_{s(i)}^x + w_i^x + e_{i,j}^x \\
y_{i,j} (t_{i,j}^y)\mid \{z_{i}=c\} =& \beta_{0c}^y  + \beta_{0,base}^y Baseline^y  + \beta_{1c}^y  t_{i,j}^y + \beta_{2c}^y (t_{i,j}^y)^2 + v_{s(i)}^y + w_i^y + e_{i,j}^y 
\end{align*}
where $v_{s(i)}^x \sim N(0, 1/\tau_{sc}^x ),v_{s(i)}^y \sim N(0, 1/\tau_{sc}^y)$ are class specific errors, $w_i^x \sim N(0, 1/\tau_w^x), $ and $w_i^y \sim N(0,1/\tau_w^y)$ are errors associated to the outcome measure, and $ e_{i,j}^x\sim N(0,1/\tau_e^x)$ and $ e_{i,j}^y\sim N(0,1/\tau_e^y)$ are common errors.
For latent class assignment, assume
\begin{align*}
z_i\sim &\ \text{Categorical}(\pi_1,\cdots,\pi_C)\\
\pi_1,\cdots,\pi_C \sim &\ \text{Dirichlet}(\alpha/C,\cdots,\alpha/C).
\end{align*} In this context, $\alpha$ is a hyperparameter set to be $\alpha\sim\text{Unif}(1,3)$, since JAGS is used for computation and can encounter numerical issues with small values of $\alpha$.
Enabling site for modification of class assignment through data augmentation, define
\begin{align*}
s_i \mid \{z_{i}=c\} \sim &\ \text{Categorical}(p_{1c},\cdots,p_{Mc})\\
p_{1c},\cdots,p_{Mc} \sim &\ \text{Dirichlet}(1,\cdots,1).
\end{align*}
Below presents the prior distributions for model parameters specified as non-informative conjugate priors for each class $c=1,\cdots, C$:
\begin{align*}
    \beta_{0c}^x&\sim N(0, 1/\tau_0^x), \beta_{1c}^x\sim N(0, 1/\tau_1^x), \beta_{1c}^x\sim N(0, 1/\tau_2^x), 1/\tau_{sc}^x \sim \text{Ga}(\gamma_{sc}^x, \gamma_{sc}^x)\\
    \beta_{0c}^y&\sim N(0, 1/\tau_0^y), \beta_{1c}^y\sim N(0, 1/\tau_1^y), \beta_{1c}^y\sim N(0, 1/\tau_2^y), 1/\tau_{sc}^y \sim \text{Ga}(\gamma_{sc}^y, \gamma_{sc}^y).
\end{align*}
Here $\text{Ga}$ is the Gamma distribution and $\gamma_{sc}^x$ and $\gamma_{sc}^y$ are chosen arbitrarily based on the multitude of the scale of the observations and variation. The hyperpriors for the precision parameters are conjugate as well:
\begin{align*}
    \tau_0^x &\sim \text{Ga}(\gamma_{0}^x, \gamma_{0}^x), \tau_1^x\sim \text{Ga}(\gamma_{1}^x, \gamma_{1}^x), \tau_2^x\sim \text{Ga}(\gamma_{2}^x, \gamma_{2}^x)\\
    \tau_0^y&\sim \text{Ga}(\gamma_{0}^y, \gamma_{0}^y), \tau_1^y\sim \text{Ga}(\gamma_{1}^y, \gamma_{1}^y), \tau_2^y\sim \text{Ga}(\gamma_{2}^y, \gamma_{2}^y)
\end{align*}
where $\gamma_{0}^x,\gamma_{1}^x,\gamma_{2}^x,\gamma_{0}^y,\gamma_{1}^y,\gamma_{2}^y$ can be set up arbitrarily. In addition, for the common parameters:
\begin{align*}
    \beta_{0,\text{base}}^x &\sim N(\mu_{0,\text{base}}^x, 1/\tau_{0,\text{base}}^x), \tau_{0,\text{base}}^x \sim \text{Ga}(\gamma_{0,\text{base}}^x, \gamma_{0,\text{base}}^x), \tau_w^x \sim \text{Ga}(\gamma_{w}^x, \gamma_{w}^x), \tau_e^x \sim \text{Ga}(\gamma_{e}^x, \gamma_{e}^x)\\
    \beta_{0,\text{base}}^y &\sim N(\mu_{0,\text{base}}^y, 1/\tau_{0,\text{base}}^y), \tau_{0,\text{base}}^y \sim \text{Ga}(\gamma_{0,\text{base}}^y, \gamma_{0,\text{base}}^y), \tau_w^y \sim \text{Ga}(\gamma_{w}^y, \gamma_{w}^y), \tau_e^y \sim \text{Ga}(\gamma_{e}^y, \gamma_{e}^y)
\end{align*}
where $\gamma_{w}^x,\gamma_{e}^x,\mu_{0,\text{base}}^x,\gamma_{0,\text{base}}^x,\gamma_{w}^y,\gamma_{e}^y,\mu_{0,\text{base}}^y, \gamma_{0,\text{base}}^y$ can be set up arbitrarily for being noninformative.

\newpage
\section{Algorithms details}
\label{alg:details}

\begin{algorithm}[hbt!]
\caption{Branching out quickly}
\begin{algorithmic}[1]
\For {steps $s=1,\cdots$ till $p_\text{sum}>c$ to branch out quickly}
\State {Select the immediate neighbours $(u_i^s+1,v_i^s), (u_i^s-1,v_i^s),(u_i^s,v_i^s+1), (u_i^s,v_i^s-1)$ for all $i\in \mathcal{M}$ and save them into a set $\mathcal{M}_1$ and get the boundary $\mathcal{D}_1$ of $\mathcal{M}_1$.}
\State { Select the next neighbors to immediate neighbours $(u_j^s+1,v_j^s), (u_j^s-1,v_j^s),(u_j^s,v_j^s+1), (u_j^s,v_j^s-1)$ for all $j\in \mathcal{M}_1$ and save them into a set $\mathcal{M}_2$. Get the boundary $\mathcal{D}_2$ of $\mathcal{M}_2$.}
\State {Order the corresponding probabilities of $\mathcal{D}_1$ from smallest to largest and rearrange the set accordingly into $\mathcal{D}_1^*$. Compute the cumulative sum of ordered probabilities $p_{1,\text{cum}}^*$. }
\State {Order the corresponding probabilities of $\{\mathcal{D}_2,\mathcal{M}_R\}$ from largest to smallest into $p_{2}^*$ and arrange the set accordingly into $\mathcal{D}_2^*$.}
\If{$\exists r \in \mathbb{N}^+$ s.t. $ \text{max}_r [p_{2}^*]_1>[p_{c1}^*]_r$}
\State  {Select the highest probability region in $\mathcal{D}_2^*$ to cover the small probability regions in $\mathcal{D}_1^*$ as many as possible, i.e. $\text{max}_r [p_{2}^*]_1>[p_{c1}^*]_r$ where $[\cdot]_r$ denotes $r$th element in the vector.} 
        \State {Remove the first $r$ regions in $\mathcal{D}_1^*$ and save it into $\mathcal{D}_1^{**}:=[\mathcal{D}_1^{*}]_{-1:r}$. Update $M=\{\mathcal{M},\mathcal{D}_1^{**}, [\mathcal{D}_2^*]_1\}$. Update the removed region set $\mathcal{M}_R=\{[\mathcal{M}_R]_{-1}, [\mathcal{D}_1^{*}]_{-1:r}\}$ if the highest probability region is from the removed region set or $\mathcal{M}_R=\{[\mathcal{M}_R], [\mathcal{D}_1^{*}]_{-1:r}\}$ if the highest probability region is from the next immediate neighbors $\mathcal{D}_2$.  } 
\Else
    \State { Otherwise, update $\mathcal{M}=\{\mathcal{M},\mathcal{D}_1\}$.}
\EndIf
\State \parbox[t]{\dimexpr\textwidth-\leftmargin-\labelsep-\labelwidth}{ Compute the sum of the posterior prediction probability $p_\text{sum}$ of all the selected regions.}
\EndFor
\end{algorithmic}
\end{algorithm}

\begin{algorithm}[hbt!]
\caption{Branching out slowly}
\begin{algorithmic}[1]
\For {Steps $s=1,\cdots$ till $p_\text{sum}>\text{target coverage}$ to slowly branch out and fine tune}
   \State { Select the boundary regions for all $i\in \mathcal{M}$ and save them into a set $\mathcal{D}_1$.}
   \State { Select the immediate neighbours $(u_j^s+1,v_j^s), (u_j^s-1,v_j^s),(u_j^s,v_j^s+1), (u_j^s,v_j^s-1)$ for all $j\in \mathcal{M}$ and save them into a set $\mathcal{D}_2$.}
   \State { Order the corresponding probabilities of $\mathcal{D}_1$ from smallest to largest and rearrange the set accordingly into $\mathcal{D}_1^*$. Compute the cumulative sum of ordered probabilities $p_{1,\text{cum}}^*$.} 
   \State { Order the corresponding probabilities of $\{\mathcal{D}_2,\mathcal{M}_R\}$ from largest to smallest into $p_{2}^*$ and arrange the set accordingly into $\mathcal{D}_2^*$.}
   \If{$\exists r \in \mathbb{N}^+$ s.t. $\text{max}_r [p_{2}^*]_1>[p_{c1}^*]_r$}
   \State { Select the highest probability region in $\mathcal{D}_2^*$ to cover the small probability regions in $\mathcal{D}_1^*$ as many as possible, i.e. $\text{max}_r [p_{2}^*]_1>[p_{c1}^*]_r$ where $[\cdot]_r$ denotes $r$th element in the vector.}
    \State { Remove the first $r$ regions in $\mathcal{D}_1^*$ and save it into $\mathcal{D}_1^{**}:=[\mathcal{D}_1^{*}]_{-1:r}$. Update $M=\{\mathcal{M},\mathcal{D}_1^{**}, [\mathcal{D}_2^*]_1\}$. Update the removed region set $\mathcal{M}_R=\{[\mathcal{M}_R]_{-1}, [\mathcal{D}_1^{*}]_{-1:r}\}$ if the highest probability region is from the removed region set or $\mathcal{M}_R=\{[\mathcal{M}_R], [\mathcal{D}_1^{*}]_{-1:r}\}$ if the highest probability region is from the next immediate neighbors $\mathcal{D}_2$.}
    \Else
        \State {Otherwise, branch out and update $\mathcal{M}=\{\mathcal{M},[\mathcal{D}_2^{*}]_1\}$.}
    \EndIf
    \State {Compute the sum of the posterior prediction probability $p_\text{sum}$ of all the selected regions.}
\EndFor
\end{algorithmic}
\end{algorithm}

\newpage
\section{Additional figures}

\begin{figure}[!htp]
    \includegraphics[width=0.5\textwidth]{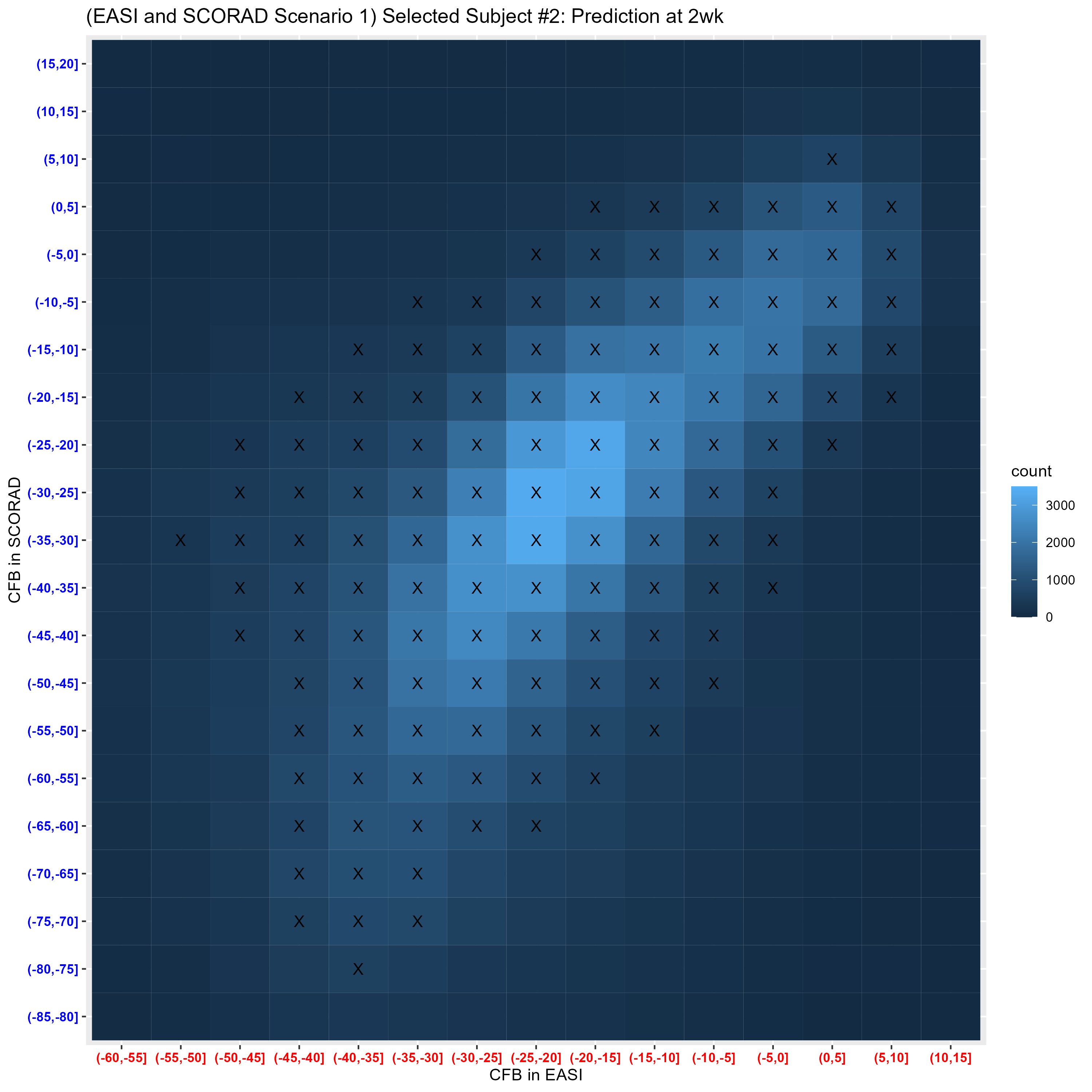}
    \includegraphics[width=0.5\textwidth]{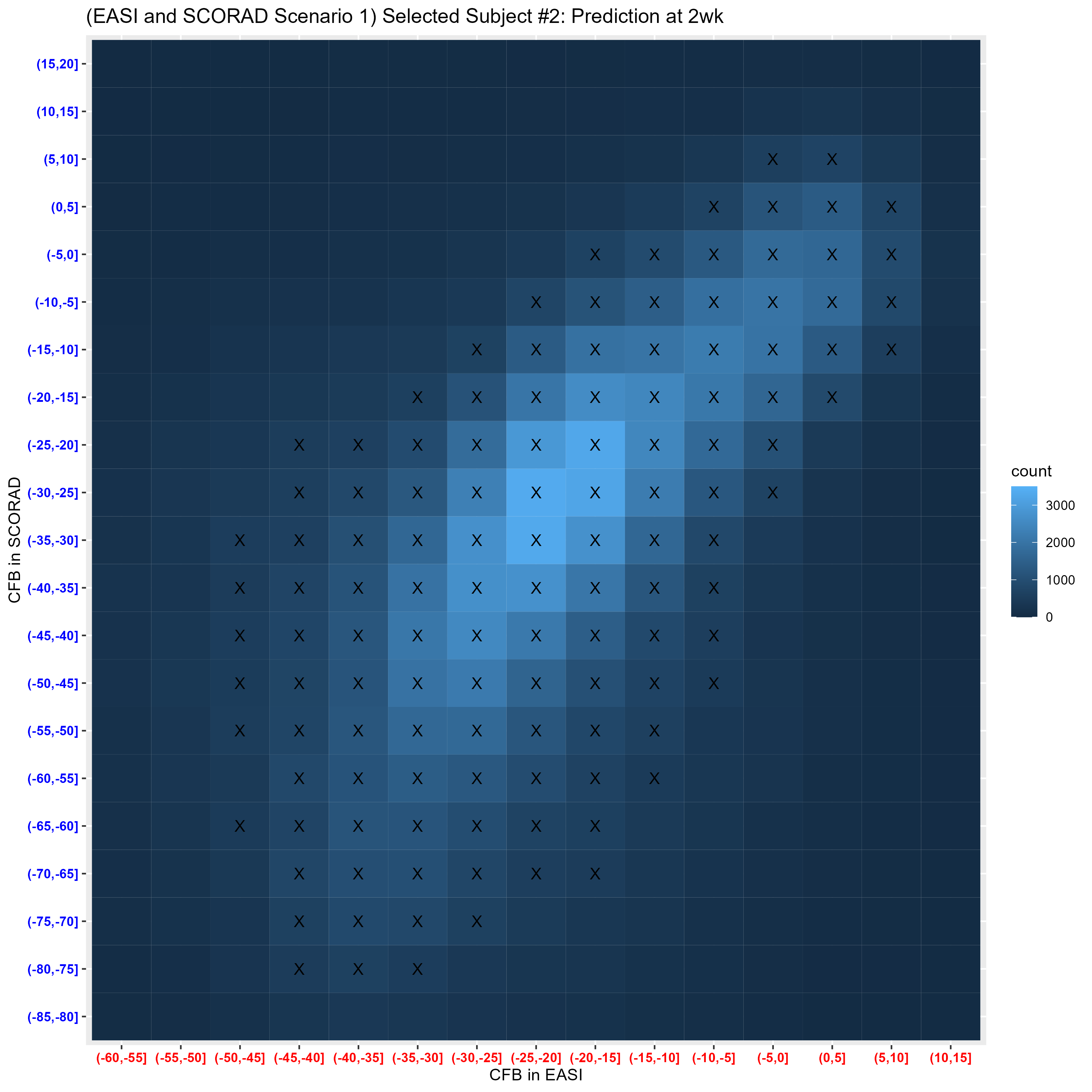}
    \caption{Prediction Scenario 1 (EASI and SCORAD): $80\%$ credible regions for Scenario 1 Selected Subject $\#2$ at Week 2 conditional on baseline. The $80\%$ credible regions are identified with X over the regions. Left: Algorithm 1 (Branching out); Right: Algorithm 2: HDR. Baseline (EASI:34.8, SCORAD:70.4); True value at Week 2 (EASI:-7, SCORAD:-15.6). The observation at Week 2 is contained in the credible regions for both algorithms.}
    \label{fig:app2pp1subj2}
\end{figure}

\begin{figure}[!htp]
    \includegraphics[width=0.5\textwidth]{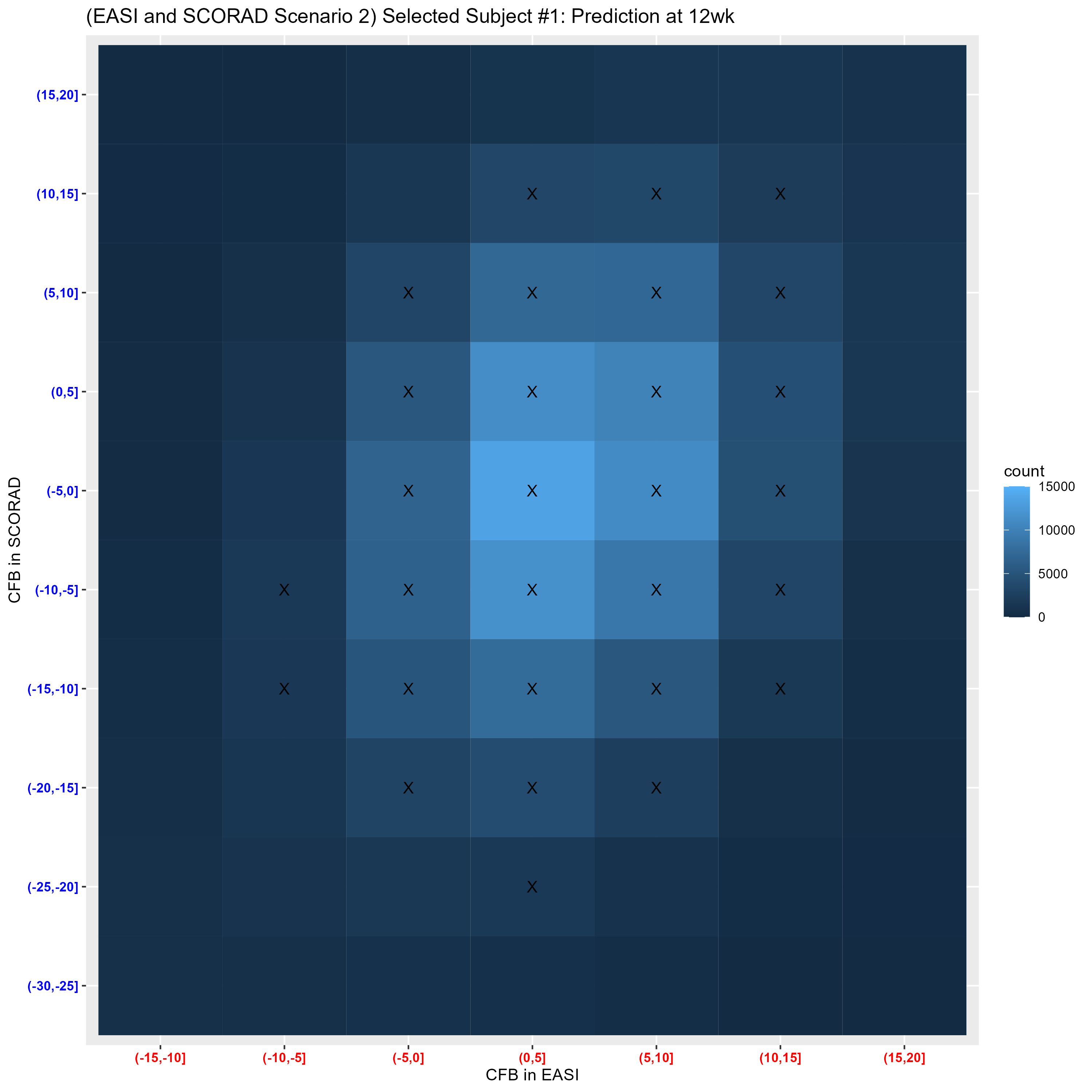}
    \includegraphics[width=0.5\textwidth]{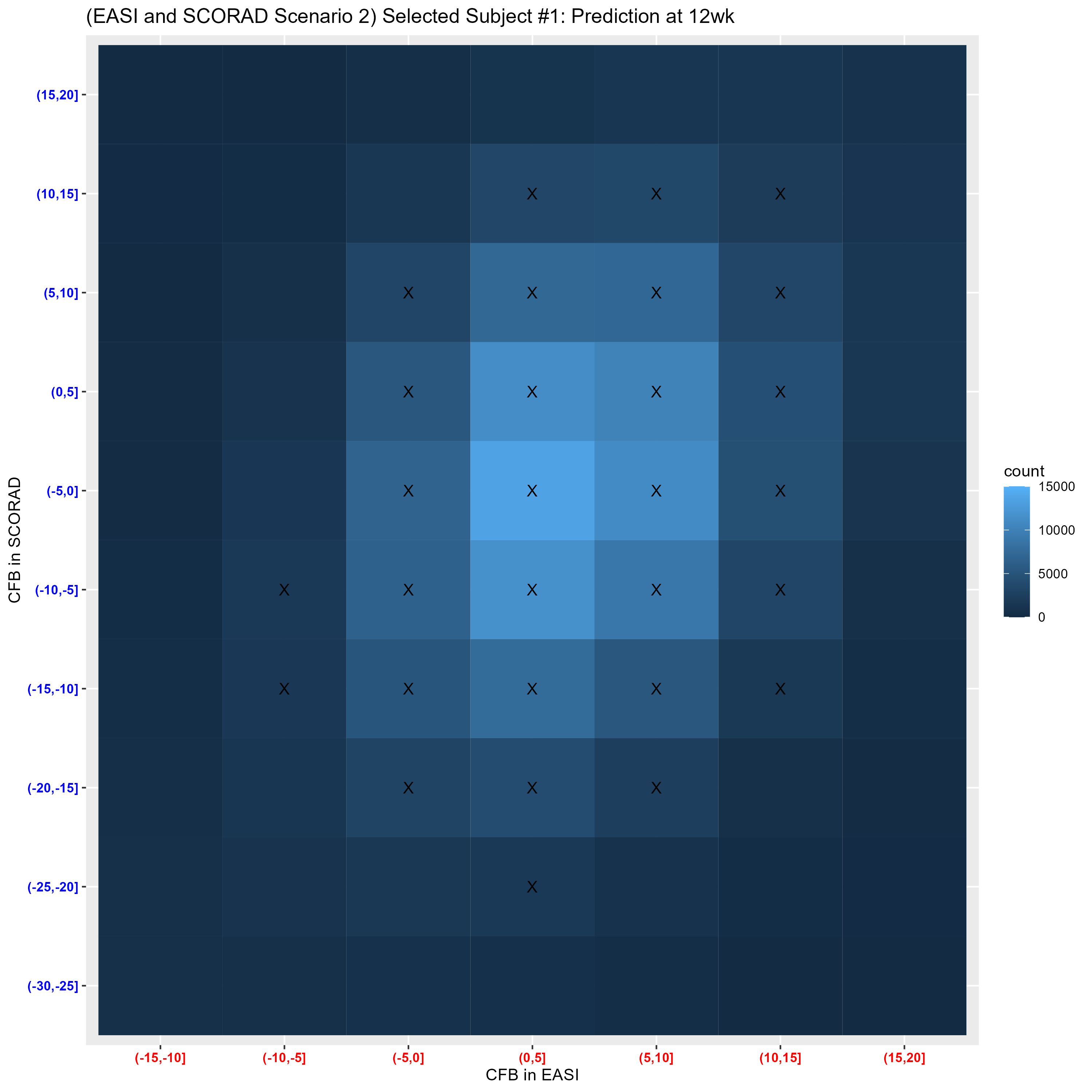}
    \caption{Prediction Scenario 2 (EASI and SCORAD): $80\%$ credible regions for Scenario 2 Selected Subject $\#1$ at Week 12 conditional on Weeks 2, 4, and 8. The $80\%$ credible regions are identified with X over the regions. Left: Algorithm 1 (Branching out); Right: Algorithm 2: HDR. Baseline (EASI:17.6, SCORAD:54.7); True value at Week 12 (EASI:-14, SCORAD:-27.2). The observation at Week 12 is not contained in the credible regions for both algorithms.}
    \label{fig:app2pp2subj1}
\end{figure}


\begin{figure}[!htp]
    \includegraphics[width=0.5\textwidth]{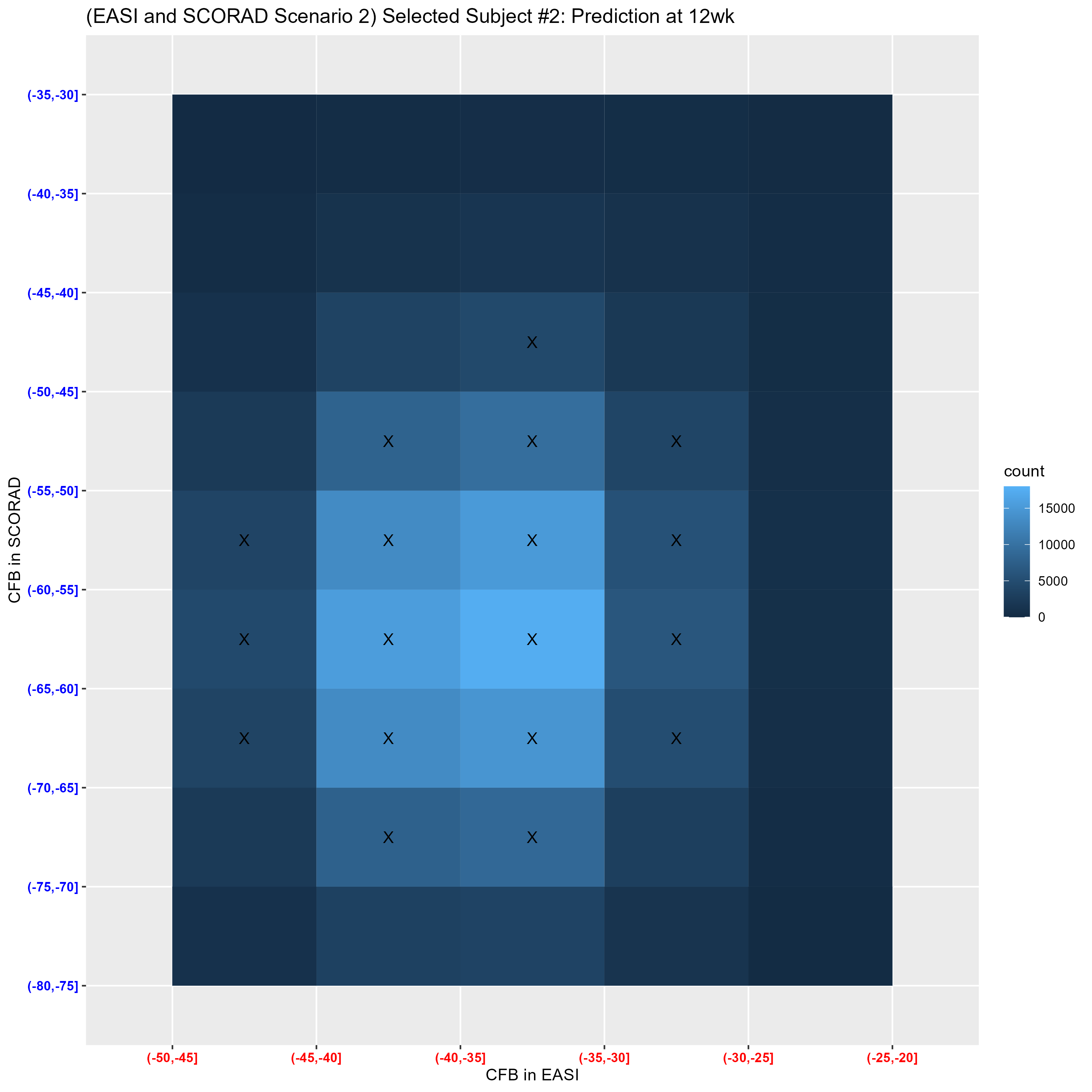}
    \includegraphics[width=0.5\textwidth]{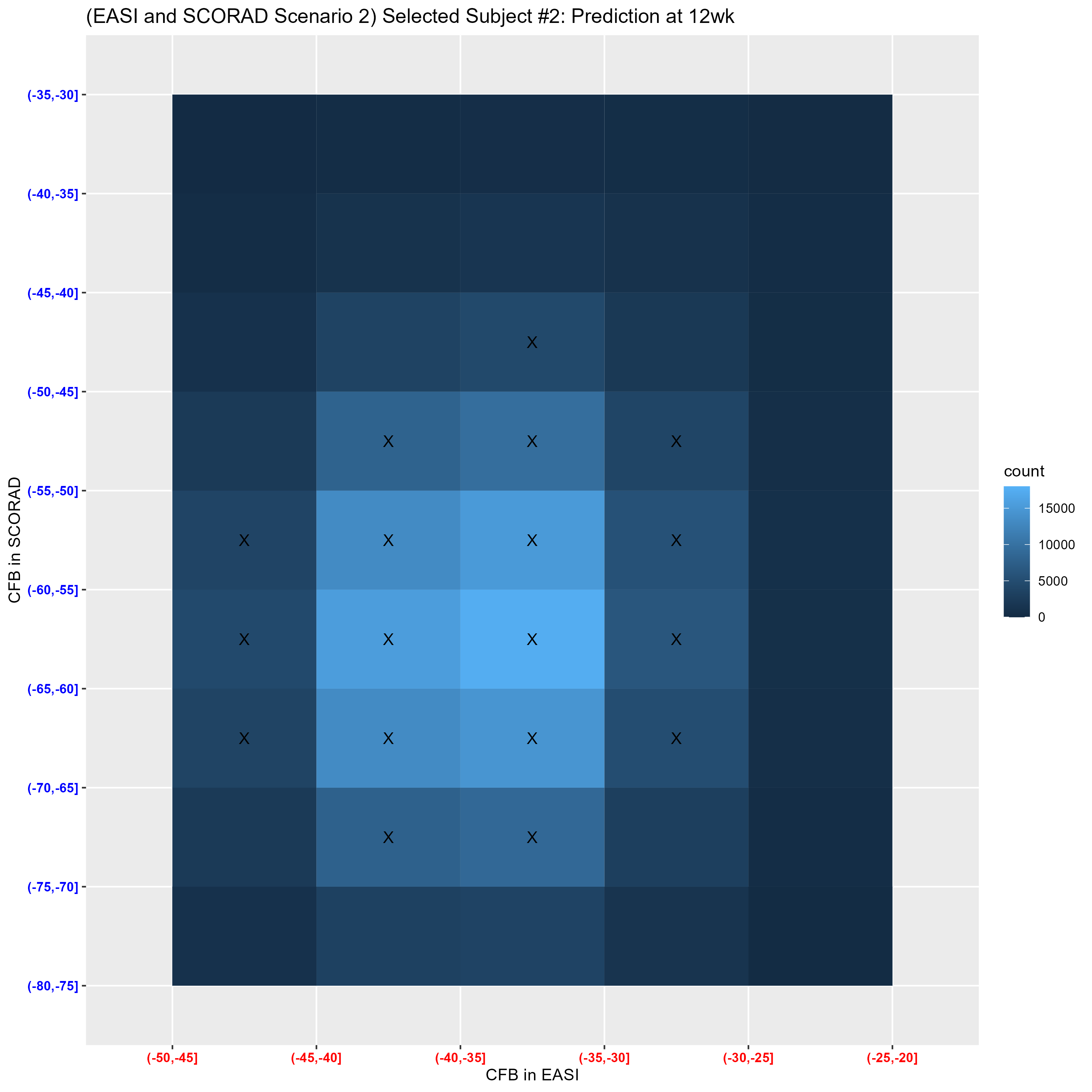}
    \caption{Prediction Scenario 2 (EASI and SCORAD): $80\%$ credible regions for Scenario 2 Selected Subject $\#2$ at Week 12 conditional on Weeks 2, 4, and 8. The $80\%$ credible regions are identified with X over the regions. Left: Algorithm 1 (Branching out); Right: Algorithm 2: HDR. Baseline (EASI:34.8, SCORAD:70.4); True value at Week 12 (EASI:-33.8, SCORAD:-55.4). The observation at Week 12 is contained in the credible regions for both algorithms.}
    \label{fig:app2pp2subj2}
\end{figure}

\begin{figure}[!htp]
    \includegraphics[width=0.5\textwidth]{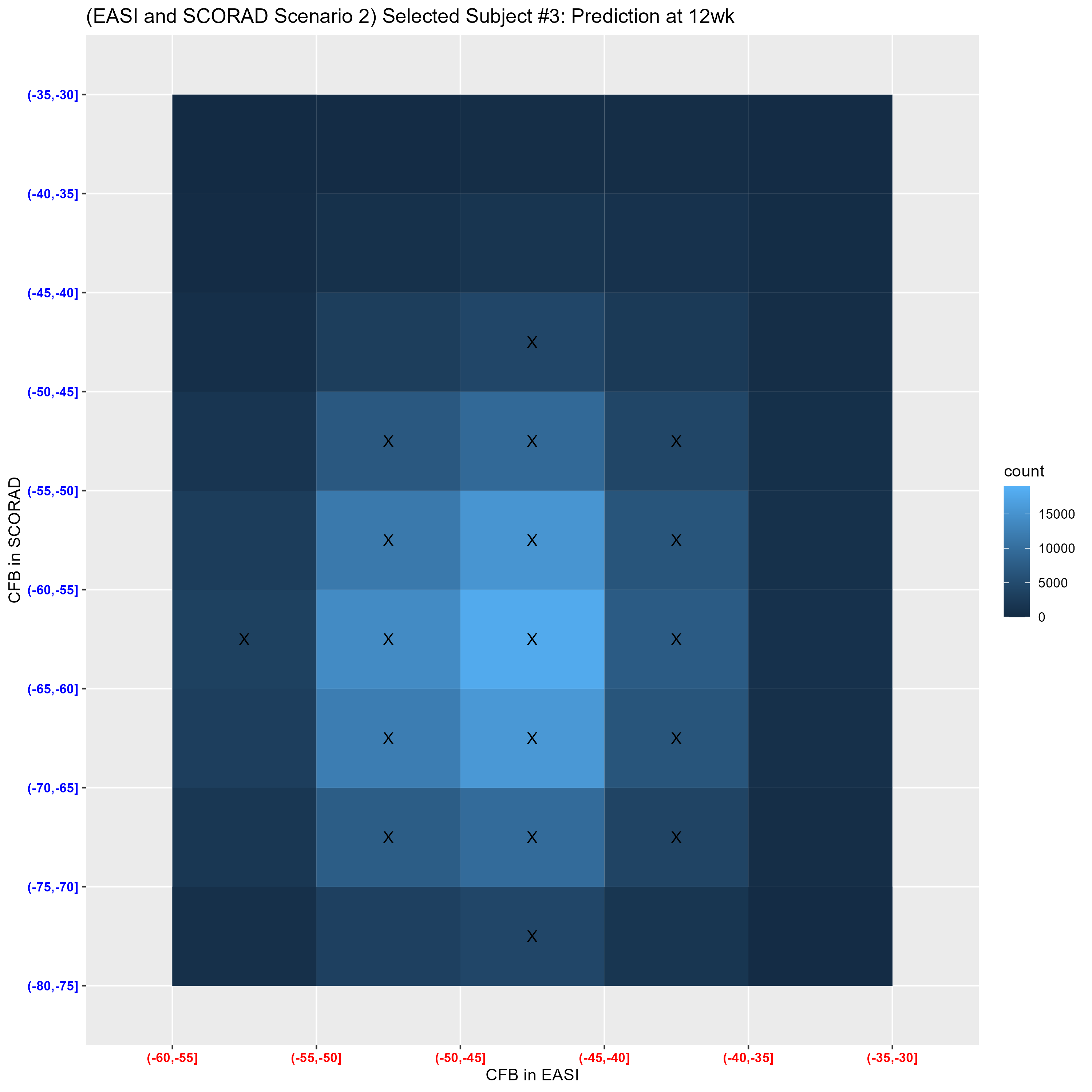}
    \includegraphics[width=0.5\textwidth]{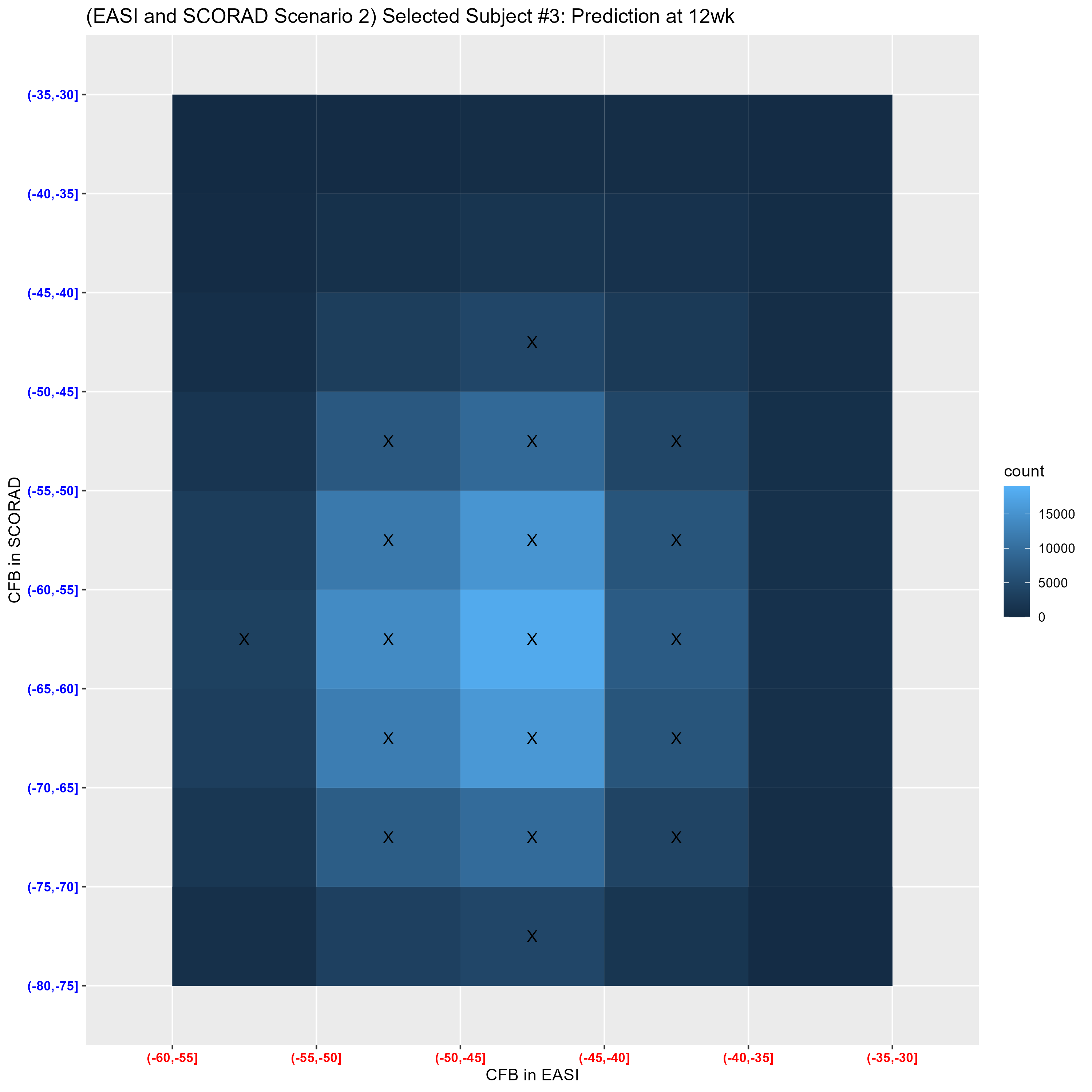}
    \caption{Prediction Scenario 2 (EASI and SCORAD): $80\%$ credible regions for Scenario 2 Selected Subject $\#3$ at Week 12 conditional on Weeks 2, 4, and 8. The $80\%$ credible regions are identified with X over the regions. Left: Algorithm 1 (Branching out); Right: Algorithm 2: HDR. Baseline (EASI:45.3, SCORAD:73); True value at Week 12 (EASI:-39.6, SCORAD:-42.9). The observation at Week 12 is not contained in the credible regions for both algorithms.}
    \label{fig:app2pp2subj3}
\end{figure}

\end{document}